%% file: ms.tex
\documentclass[conference]{IEEEtran}
\usepackage{cite} %[noadjust]
\usepackage{amsmath,amssymb,amsfonts}
\usepackage{algorithmic}
\usepackage{graphicx}
\usepackage{textcomp}
\usepackage{bmpsize}
\usepackage[colorlinks=true,urlcolor=black]{hyperref}
\def\BibTeX{{\rm B\kern-.05em{\sc i\kern-.025em b}\kern-.08em
    T\kern-.1667em\lower.7ex\hbox{E}\kern-.125emX}}

\usepackage{cleveref}
\crefformat{section}{\S#2#1#3} % see manual of cleveref, section 8.2.1
\crefformat{subsection}{\S#2#1#3}
\crefformat{subsubsection}{\S#2#1#3}
\crefformat{appendix}{\S#2#1#3}

\usepackage[svgnames]{xcolor,colortbl} % to color table columns
\usepackage{enumitem}
\usepackage{adjustbox}
\usepackage{subcaption}
\usepackage{graphics}
\usepackage{xcolor,colortbl}
\usepackage{multirow, hhline}
\usepackage{comment}

\definecolor{Gray1}{gray}{0.88}
\definecolor{Gray2}{gray}{0.93}
\definecolor{Gray3}{gray}{0.975}
\definecolor{LightCyan}{rgb}{0.88,1,1}

\newcolumntype{a}{>{\columncolor{Gray1}}c}
\newcolumntype{b}{>{\columncolor{Gray2}}c}
\newcolumntype{d}{>{\columncolor{Gray3}}c}

\newcommand{\impvm}[2]{${#1}\textsc{VM-imp}_{\textit{#2}}$}
\newcommand{\normvm}[1]{${#1}\textsc{VM}$}
\newcommand{\ropobf}[1]{$\textsc{ROP}_{#1}$}
\newcommand{\ropother}[1]{$\textsc{ROP-P}_{#1}$}

\usepackage{etoolbox}
\makeatletter
\patchcmd{\@verbatim}
  {\verbatim@font}
  {\verbatim@font\footnotesize}
  {}{}
\makeatother

\usepackage{xspace}
\newcommand{\goalone}{\textsf{G\textsubscript{1}}\xspace}
\newcommand{\goaltwo}{\textsf{G\textsubscript{2}}\xspace}
\newcommand{\goalboth}{\textsf{G\textsubscript{1-2}}\xspace}
\newcommand{\surfaceone}{\textsf{A\textsubscript{1}}\xspace}
\newcommand{\surfacetwo}{\textsf{A\textsubscript{2}}\xspace}
\newcommand{\surfacethree}{\textsf{A\textsubscript{3}}\xspace}
\newcommand{\surfacesome}[1]{\textsf{A\textsubscript{#1}}\xspace}
\newcommand{\cse}{\textsf{SE}\xspace}
\newcommand{\dse}{\textsf{DSE}\xspace}
\newcommand{\tds}{\textsf{TDS}\xspace}
\newcommand{\predone}{\textsf{P\textsubscript{1}}\xspace}
\newcommand{\predtwo}{\textsf{P\textsubscript{2}}\xspace}
\newcommand{\predthree}{\textsf{P\textsubscript{3}}\xspace}
\newcommand{\predsome}[1]{\textsf{P\textsubscript{#1}}\xspace}

\newcommand{\scz}[1]{{\color{black} #1}} % blue
\newcommand{\puh}[1]{{#1}\xspace}%{\color{gray} #1} \color{red}

\newif\ifonlinereport
\onlinereporttrue
\ifonlinereport
\newcommand{\apxref}{\textsection\Cref{apx:a}\xspace} % only for quick reference
\else
\IEEEtriggeratref{76}
\newcommand{\apxref}{\cite{our-extended}\xspace}
\fi

\IEEEoverridecommandlockouts

\begin{document}

\title{Hiding in the Particles: When Return-Oriented Programming Meets Program Obfuscation{*\thanks{* Online extended version for the paper published in the Proceedings of DSN'21 (51st IEEE/IFIP Int. Conf. on Dependable Systems and Networks). Code and BibTeX entry available at \url{https://github.com/pietroborrello/raindrop}.}}}

\author{\IEEEauthorblockN{Pietro Borrello}%
\IEEEauthorblockA{\textit{Sapienza University of Rome} \\%
%City, Country \\
borrello@diag.uniroma1.it}%
\and%
\IEEEauthorblockN{Emilio Coppa}%
\IEEEauthorblockA{\textit{Sapienza University of Rome} \\%
%City, Country \\
coppa@diag.uniroma1.it}%
\and%
\IEEEauthorblockN{Daniele Cono D'Elia}%
\IEEEauthorblockA{\textit{Sapienza University of Rome} \\%
delia@diag.uniroma1.it}}

\maketitle

\ifonlinereport
\thispagestyle{plain}
\pagestyle{plain}
\fi

\begin{abstract}
Largely known for attack scenarios, code reuse techniques at a closer look reveal properties that are appealing also for program obfuscation. We explore the popular return-oriented programming paradigm under this light, transforming program functions into ROP chains that coexist seamlessly with the surrounding software stack. We show how to build chains that can withstand popular static and dynamic deobfuscation approaches, evaluating the robustness and overheads of the design over common programs. The results suggest a significant amount of computational resources would be required to carry a deobfuscation attack for secret finding and code coverage goals.
\end{abstract}


\input{intro}
\input{background}

\input{adversarial-model}

\input{design}

\input{strengthening-rop}
\input{related}

\input{evaluation}

\input{discussion}
\input{conclusion}

\bibliographystyle{IEEEtran}
\bibliography{biblio}

\ifonlinereport
\input{appendix}

\fi

\end{document}

%% file: intro.tex
% !TEX root = main.tex
% !TEX root = main.tex

\section{Introduction}
\label{se:intro}

%Ciao DSN!

Memory errors are historically among the most abused software vulnerabilities for arbitrary code execution exploits~\cite{vanderVeen-RAID12}. Since the introduction of system defenses against code injection attempts, code reuse techniques earned the spotlight for their ability in reassembling existing code fragments of a program to build the execution sequence an attacker desires.%achieve the result an attacker desires.

% with Jekyll programs turning evil at an attacker's command
%  and signing
{\em Return-oriented programming} (ROP)~\cite{Shacham07} is the most eminent code reuse technique. \puh{Thanks} to its rich expressivity,  \puh{ROP has also} seen several uses besides exploitation. Researchers have used \puh{it constructively, for instance, in code integrity verification~\cite{parallax},} or maliciously to embed hidden functionality in code that undergoes auditing~\cite{jekyll,ropneedle}. Security firms have reported cases of malware in the wild written in ROP~\cite{fireeye666}.

% target standard instruction pointer-driven code
Some literature considers ROP code bothersome to analyze: humans may struggle with the exoticism of the representation, and the vast majority of tools used for code understanding and reverse engineering \puh{have no provisions for code reuse payloads}~\cite{DebraySP15,synthia,ropmemu,ropscozzo}. Automatic proposals for analyzing complex ROP code started to emerge only \puh{recently}~\cite{DebraySP15, ropmemu, ropscozzo}.%in recent years

\puh{We believe that the quirks of the ROP paradigm offer promising opportunities to realize effective code obfuscation schemes. In this paper we present a protection mechanism that builds on ROP to hide implementation details of a program from motivated attackers that can resort to a plethora of automated code analyses. We analyze what qualities make ROP appealing for obfuscation, and address its weak links to make it robust in the face of an adversary that can symbiotically combine general and ROP-aware code analysis methods.}
%We came to believe that the quirks of the ROP paradigm offer promising opportunities for software protection research, turning ROP into a defensive mechanism that can hide the implementation details of a program from motivated attackers that can resort to a plethora of automated code analyses~\cite{SurveyCSUR16}. In this paper we analyze what qualities make ROP appealing for {\em program obfuscation}, and address its weak links to make it robust in the face of an adversary that can symbiotically combine classic and ROP-aware code analysis methods. 

\subsubsection*{Motivation}
From a code analysis perspective, we observe that the control flow of a ROP sequence is naturally destructured. Each ROP gadget ends with a {\tt ret} instruction that operates like a dispatcher in a language interpreter: {\tt ret} reads from the top of the stack the address of the next gadget and transfers control to it. The stack pointer RSP becomes a {\em virtual program counter} for the execution, sidelining the standard instruction pointer RIP, while gadget addresses become the instructions supported by this custom language.

% for any
\looseness=-1% MAGIA PURA
This level of {\em indirection} makes the identification of basic blocks and of control transfers between them not immediate. This challenges humans and classic disassembly and decompilation approaches, but may not be an issue for dynamic deobfuscation approaches that explore the program state space systematically (e.g., symbolic execution~\cite{survey-baldoni}) or try to extricate the original control flow from the dispatching logic (e.g.,~\cite{DebraySP15}), nor for ROP-aware analyses that dissect RSP and RIP changes. Protecting transfers is critical for  \puh{program obfuscations} to withstand advanced deobfuscation methods, and we introduce three ROP transformations that address this weak link.
%which as we will see is also a recurrent issue in existing protection schemes.

Another benefit from using ROP for obfuscation is the {\em code diversity}~\cite{franz} it can bring. \puh{Obfuscations may randomize the instructions emitted at specific points, but can incur a limited transformation space~\cite{TutorialBanescu}.} We can use multiple equivalent gadgets in the encoding to serve one same purpose in different program points. But one gadget can also serve different purposes in different points: the instructions in it that concur to the program semantics will depend on the surrounding chain portion, while the others are dynamically dead. This not only complicates manual analysis, but helps also against pattern attacks that may try to recognize specific gadget sequences to deem the location of ROP branches and blocks in the chain.

%Pattern attacks are also important as they can complement an attacker's toolbox~\cite{bardin-acsac19}. With standard code
% (i.e., opcode bytes)
% introduce
\puh{Such attacks often complement} an attacker's toolbox~\cite{bardin-acsac19}: for instance, an adversary may heuristically look for distinctive instructions in memory and try to patch away parts that hinder \puh{semantic attacks}. We identify a distinctive benefit of ROP: the adversary only sees bytes that form gadget addresses or data operands, and because of indirection needs to dereference addresses to retrieve the actual instructions. With a careful encoding we can induce \puh{{\em gadget confusion}} that makes it harder \puh{also to locate the position of gadget addresses in the chain}.%\puh{distinguish data from code (i.e. gadget addresses)}.

% locate bytes that represent gadget addresses or data.%, raising the bar against pattern attacks as well.

\subsubsection*{Contributions}
%In this paper we bring novel ideas to the software protection realm, broadening the horizon of previous research that explored constructive uses of ROP. We present and implement a design to transform in their entirety program functions into ROP chains that interact seamlessly with standard software stack components. We shield chains using natural encoding extensions that raise the bar for generic classes of deobfuscation attacks, and evaluate 72 standard synthetic functions in two common attack scenarios, putting the deobfuscation computational effort into perspective with several configurations of \puh{the prominent} virtualization obfuscation~\cite{TutorialBanescu}. We also study the slowdowns of our techniques on performance-sensitive code, and evaluate their coverage on a heterogeneous real-world code base. \puh{In summary, over} the next sections we present:
\puh{In this work we bring novel ideas to the software protection realm, presenting a protection mechanism that significantly slows down or deters current automated deobfuscation attacks. We show how to transform entire program functions into ROP chains that interact seamlessly with standard code components, introducing novel natural encoding transformations that raise the bar for general classes of attacks. We evaluate our techniques over synthetic functions for two common deobfuscation tasks, putting the computational effort for succeeding into perspective with different configurations of the prominent~virtualization obfuscation~\cite{TutorialBanescu}. We also analyze their slowdowns on performance-sensitive code, and their coverage on a heterogeneous real-world code base. In summary, over} the next sections we present:
\begin{itemize}
\item a rewriter \puh{that turns compiled functions into ROP chains};
%a ROP rewriter to transform binary functions into chains;
% three generic classes of deobfuscation attacks
\item an analysis of ROP in the face of three attack surfaces for general deobufscation, and three encoding \puh{predicates} that increase \puh{the} resistance against such attacks;
\item \puh{a resistance study for secret finding and code coverage goals with symbolic, taint-driven, and ROP-aware tools;}
\item a coverage study where we transform $95.1\%$ of the unique functions \puh{composing} the {\tt coreutils} Linux suite.
\end{itemize}

% research for, rewriter
In the hope of fostering further \puh{work in} program protection, we make our \puh{system} available to researchers.
Details for access can be found at \url{https://github.com/pietroborrello/raindrop/}.

%%%% END of file

%% file: background.tex
% !TEX root = main.tex

\section{Preliminaries}
\label{se:background}

This section details key concepts from code obfuscation and ROP research that are relevant to the ideas behind this paper.

%In this section, we introduce key concepts from code obfuscation and ROP research that are relevant to our proposal. We then present the adversarial model considered throughout the paper, describing also state-of-the-art techniques that could be used to dismantle obfuscation approaches.

\subsection{Code Obfuscation}
\label{ss:code-obfuscation}

Software obfuscation protects digital assets~\cite{DRM} from malicious entities that some literature identifies as MATE (man-at-the-end) attackers~\cite{TutorialBanescu}. Before a research community was even born, in the '80s these entities challenged and subverted anti-piracy schemes from vendors, and shielded their own malware.

Today it represents an active research area, with heterogeneous protection mechanisms challenged by increasingly powerful program analyses~\cite{SurveyCSUR16}. Data transformations alter the position and representation of values and variables, while code transformations affect the selection, orchestration, and arrangement of instructions. Our focus are code transformations that prevent an adversary from understanding the program logic.

% as we will see
The interpretation capabilities of an attacker can be syntactic, semantic, or both. This distinction makes a great impact: for instance, instruction substitution or the insertion of spurious computations get in the way of syntax-driven attacks, but may hardly affect a semantic interpretation \puh{as we discuss in~\Cref{se:adversarial-model}}. When facing mixed capabilities, the most resilient protection schemes are often heavy-duty transformations that deeply affect the control flow and instructions of a program. 

Such transformations commonly operate at the granularity of individual functions~\cite{TutorialBanescu}. {\em Control-flow flattening}~\cite{wang,chow} collapses all the basic blocks of the control-flow graph (CFG) into a single layer, introducing a dispatcher block that picks the next block to execute based on an augmented program state. After the successful deobfuscation attack \puh{of}~\cite{debray-udupa}, \puh{present} variants try to complicate the analysis of the dispatcher~\cite{Johansson}.

{\em Virtualization obfuscation}~\cite{TutorialBanescu} completely removes the original layout and instructions: it transforms code into instructions for a randomly generated architecture and synthesizes an interpreter for it~\cite{Kinder-WCRE12}. The instructions form a bytecode representation in memory, and the interpreter maintains a virtual program counter over it: it reads each instruction and dispatches an {\em opcode handler} function that achieves the desired semantics for it. As its working resembles a virtual machine, the transformation is also known as VM obfuscation.
%
%The interpreter reads them from memory and dispatches {\em opcode handler} functions that achieve the desired semantics for a given instruction type. The instructions thus form a bytecode representation, and the interpreter maintains a virtual program counter over it. As it resembles a virtual machine, the technique is also called VM obfuscation.
%
This technique has lately monopolized the agenda of much deobfuscation research in security conferences (e.g.,~\cite{synthia, debray-ccs15, salawan, rotalume, debray-semantics}). VM obfuscation tools have three main strengths: complex code used in opcode handlers to conceal their semantics, obfuscated virtual program counter updates, and scarce reuse of deobfuscation knowledge  as the instruction set and the code for opcode handlers are generated randomly for each program.

\looseness=-1
Best practices often use data transformations at strategic points (e.g., VM dispatcher) in the implementation of a code transformation. The most common instance are {\em opaque predicates}~\cite{Collberg97}: expressions whose outcome is independent of their constituents, but hard to determine statically for an attacker. Opaque predicates can build around mathematical formulas and conjectures, mixed boolean-arithmetic (MBA) expressions, and instances of other hard problems like {\em aliasing}~\cite{ramalingan94}.

\subsection{Return-Oriented Programming}
\label{ss:rop}

\begin{figure}[t!]
    \centering
    \includegraphics[width=0.95\columnwidth]{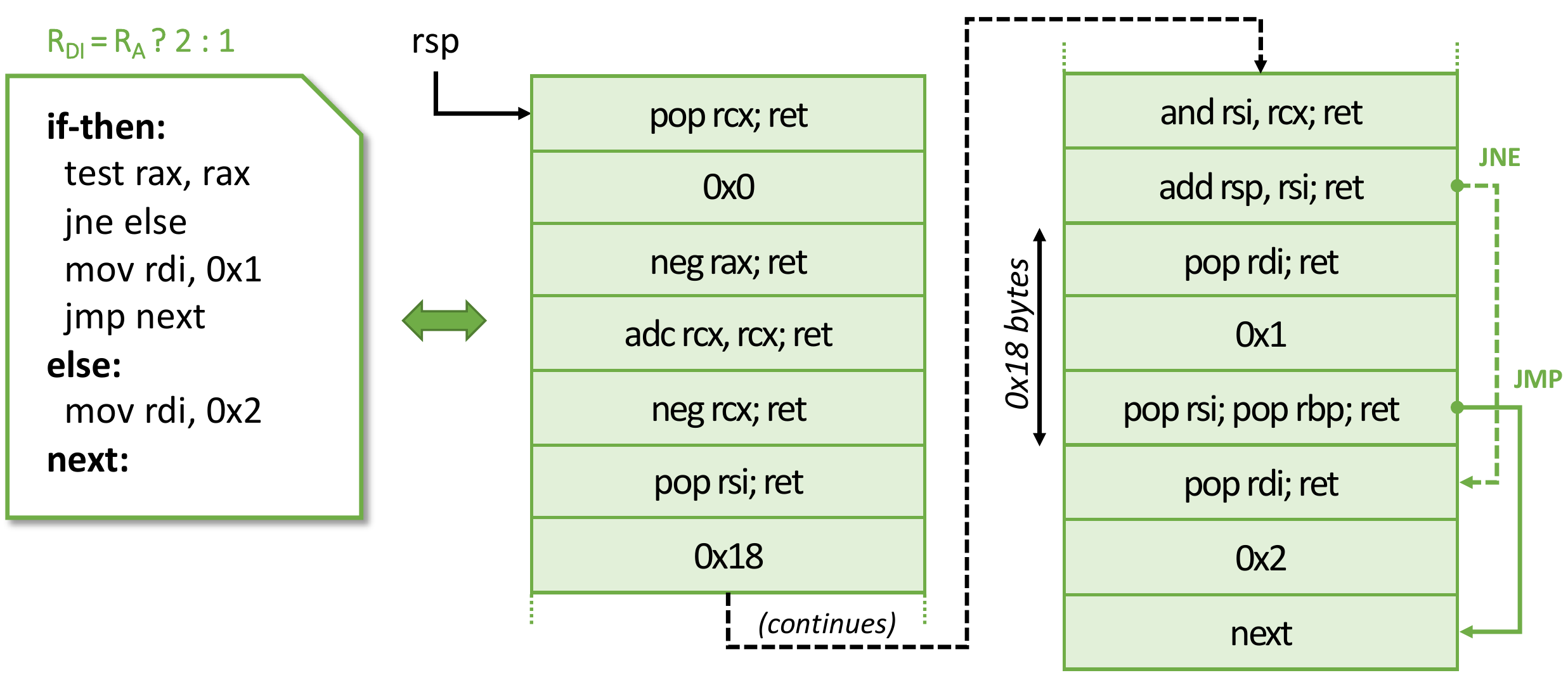}%0.95
    \vspace{-1mm}
    \caption{ROP chain with non-linear control flow. For readability pointed-to instructions appear in place of gadget addresses.}
    \label{fig:rop-branch}
    \vspace{-3mm}
\end{figure}

ROP is a technique to encode arbitrary behavior in a program by borrowing and rearranging code fragments, also called gadgets, that are already in the program~\cite{Shacham07}. Each gadget delivers a piece of the desired computation and terminates with a {\tt ret} instruction, which gives the name to the technique. A ROP payload comprises a sequence of gadget addresses interleaved with immediate data operands. The key to ignition is a pivoting sequence that hijacks the stack pointer, so that on a function return event the CPU fetches the instructions from the first gadget. Each gadget eventually transfers execution to the next using its own {\tt ret} instruction, realizing a ROP chain. %Then each gadget will transfer

% quirky aspects of the ROP practice
% is zero
\Cref{fig:rop-branch} features a chain that assigns register RDI with $1$ when register RAX==0, and with $2$ otherwise. The example showcases exoticisms of the representation with branch encoding and path-dependent semantics of chain items. The first gadget writes the immediate value {\tt 0x0} to RCX, and RSP advances by {\tt 0x10} bytes \puh{for} its {\tt pop} and {\tt ret} instructions. The next two gadgets check if RAX is zero with {\tt neg rax}: the carry flag becomes 0 when RAX==0 and 1 otherwise, then an addition with carry writes this quantity  into RCX.

ROP control-flow branches are variable RSP addends computed over a leaked CPU condition flag. The chain determines whether to skip over the {\tt 0x18} byte-long portion that sets RDI to $1$: it computes in RSI an addend that is equal to $0$ when RAX==0, and to  {\tt 0x18} otherwise, using a two's complement and a bitwise AND on RCX. If the branch is taken, RSP reaches a {\tt pop rdi} gadget that reads and assigns $2$ to RDI as desired. When execution falls through, a similar sequence sets RDI to $1$, then unconditionally jumps over the alternative assignment sequence\puh{:} this time we find no RSP addition, but a gadget disposes of the alternative $0x10$ byte-long segment by popping two junk immediates to RSI and RBP.

% , also thanks to unintended instruction alignments
Attackers can find Turing-complete sets of gadgets in mainstream software~\cite{Shacham07,exploit-hardending}. While a few works address automatic generation of ROP payloads~\cite{exploit-hardending}, publicly available tools often produce incomplete chains in real-world scenarios~\cite{vizsec18} or do not support branches. Reasons for this failure are side effects from undesired code in found gadgets, register conflicts during chaining~\cite{vizsec18}, and unavailability of ``straightforward'' gadgets for some tasks~\cite{roemerorschacham}. Albeit improved tools continue to appear (e.g.,~\cite{nrop}), no general solution for automatic ROP code generation seems available to date.

ROP is the most popular but not the sole realization of code reuse: {\tt jmp}-ended gadgets (JOP)~\cite{jop}, counterfeit C++ objects (COOP)~\cite{coop}, and other elements can be abused as well. But most importantly, ROP today is no longer only a popular mean to get around and disable code injection defenses.

Researchers and threat actors used its expressivity to create userland~\cite{fireeye666}, kernel~\cite{hund09,chuck}, and enclave~\cite{sgx-rop} malware, and to fool antivirus engines~\cite{ropinjector,ropneedle} and application review~\cite{jekyll}. The sophistication of these payloads went in some cases beyond what a human analyst can manually investigate~\cite{ropmemu}, and researchers in the meantime explored automated approaches to untangle ROP chains: we discuss these works in detail in~\Cref{se:adversarial-model}.

%Defenses against code reuse techniques include control-flow integrity (CFI) mechanisms~\cite{cfi-csur}, shadow stacks~\cite{sok-shadow}, detection of patterns at instruction~\cite{kbouncer} or micro-architectural~\cite{ropsentry} level, code randomization~\cite{ASLR} and debloating~\cite{michalis-debloating}, and others. %In response, attack vectors continue evolving, featuring for instance constraint-driven gadget discovery against CFI~\cite{newton} or data-only attacks~\cite{dop}. As those works are not directly related to the scope of our paper, we do not cover them in detail.

%Albeit the most popular, ROP is not the sole embodiment of code reuse techniques: over the years researchers explored variants based on {\tt jmp}-ended gadgets (JOP)~\cite{jop}, counterfeit C++ objects~\cite{coop}, fake signal frames~\cite{sigreturn}, and others. Defenses against code reuse techniques include control-flow integrity (CFI) mechanisms~\cite{cfi-csur}, shadow stacks~\cite{sok-shadow}, detection of patterns at instruction~\cite{kbouncer} or micro-architectural~\cite{ropsentry} level, code randomization~\cite{ASLR} and debloating~\cite{michalis-debloating}, and others. In response, attack vectors continue evolving, featuring for instance constraint-driven gadget discovery against CFI~\cite{newton} or data-only attacks~\cite{dop}. As those works are not directly related to the scope of our paper, we do not cover them in detail.

%% file: adversarial-model.tex
% !TEX root = main.tex

\section{Adversarial Model}
\label{se:adversarial-model}

This paper considers a motivated and \puh{experienced} attacker that can examine a program both statically and dynamically. The attacker is aware of the design of the used obfuscation, but not of the obfuscation-time choices made when instantiating the approach over a specific program to be protected (e.g. at which program locations we applied some transformation).

While the ultimate end goal of a reverse engineering attempt can be disparate, we follow prior deobfuscation literature (e.g.,~\cite{banescu-acsac16,Banescu-USEC17,bardin-acsac19}) in considering two deobfuscation goals that are sufficiently generic and analytically measurable:
\begin{description}[labelindent=0.5em, leftmargin=2.2em]
\item[{\goalone}\,\,\,Secret finding.] The program performs a complex computation on the input, such as a license key validation, and the attacker wishes to guess the correct value;
\item[{\goaltwo}\,\,\,Code coverage.] \puh{The attacker exercises enough (obfuscated) paths to cover all reachable (original) program code, e.g. to later analyze execution traces.}
%The attacker wishes to extract all the paths that lead to discover new code portions, for instance to later simplify them and attempt code understanding.
\end{description}

% publicly available embodiments for them
The attacker has access to state-of-the-art systems suitable for automated deobfuscation and can attempt to {\em symbiotically} combine them, using one to ease another. In the following we describe the most powerful and promising approaches available to attackers, and enucleate three attack surfaces for general deobfuscation. Those will drive our ROP encoding techniques to build chains that \puh{may} withstand such attacks. % can

% define three attack surfaces for general deobfuscation that we identified, and analyze how they affect ROP code. Then we present the details of program analysis techniques that tackle them. 

\subsection{Principles behind Automated Deobfuscation}
\label{ss:principles}

Banescu et al.~\cite{banescu-acsac16} identify a common pre-requisite in automated attacks perpetrated by reverse engineers: the need for building a suite of inputs that exercise the different paths a protected program can actually take. Achieving a coverage as high as 100\% represents \goaltwo for our attacker, while for \goalone depending on the specific function fewer paths may suffice but also data dependencies should be solved. Slowing down the generation of a ``test suite'' for the attacker is a first cut of the effectiveness of an obfuscation~\cite{banescu-acsac16}, as it is a key step in most deobfuscation pipelines for utterly disparate end goals.%~\cite{DRM,SurveyCSUR16}.

By analyzing deobfuscation research, we abstract three general attack surfaces that a ``\puh{good}'' obfuscation \puh{shall} consider: %may
\begin{description}[labelindent=0.5em, leftmargin=2.2em]
\item[{\surfaceone}\,\,\,Disassembly.] It should not be immediate for an attacker to discover code portions using static analysis techniques;
\item[{\surfacetwo}\,\,\,Brute-force search.] Syntactic code manipulations such as tracking and ``inverting'' the direction of control transfers should not reveal new code, but further dependencies must be solved in order to take valid alternate paths;
%Syntactic manipulations should not reveal code portions not exercised in an execution trace, for instance by ``inverting'' the direction of recorded control transfers, because additional dependencies should be solved
\item[{\surfacethree}\,\,\,State space.] When an obfuscation makes provisions to artificially extend the program state space to be explored,  analyses based on forward and backward dependencies of program variables should fail to simplify them away.
% Common analyses for forward and backward dependencies on program inputs and outputs should not simplify away artificial expansion mechanisms for program state.
\end{description}

In the next section we present eminent approaches for such automated attacks, which we then consider in \Cref{se:evaluation} to evaluate our ROP obfuscation. How to transform an existing program function into a ROP chain and make it robust against these three attack types are the subject of \Cref{se:encoding} and \Cref{se:strengthening}, respectively.

%We will now present eminent approaches for automated attacks and discuss their interplay with these three dimensions. 

\subsection{State-of-the-art Deobfuscation Solutions}
\label{ss:deobfuscation-solutions}

\subsubsection{General Techniques}
% also on
Deobfuscation attacks can draw from static and dynamic program analyses. Several static techniques are capable of reasoning \puh{about} run-time properties of the program, and may be the only avenue when the attacker cannot readily bring execution to a protected program portion or control its inputs. In this context, {\em symbolic execution} (\cse) reveals the multiple paths a piece of code may take by making it read symbolic instead of concrete input values, and by collecting and reasoning on path constraints over the symbols at every encountered branch. Upon termination of each path, an SMT solver generates a concrete input to exercise it~\cite{survey-baldoni}.

Scalability issues often cripple static approaches, and dynamic solutions may try to get around them by leveraging facts observed in a concrete execution. {\em Dynamic symbolic execution} (\dse) interleaves concrete and symbolic execution, collecting constraints at branch decisions that are now determined by the concrete input values, and generates new inputs by negating the constraints collected for branching decisions.

% generic, on
Obfuscators can however induce constraints that are hard for a solver, or expand the program state space artificially. Building on the intuition that these transformations are not part of the original program semantics, {\em taint-driven simplification} (\tds) tracks explicit and implicit flows of values from inputs to program outputs, untangling the control flow of an obfuscation method apart from that of the original program~\cite{DebraySP15}. \tds is a \puh{general}, dynamic, and semantics-based technique: it applies a selection of semantics-preserving simplifications \puh{to} a recorded trace and produces a simplified CFG. \tds can operate symbiotically with \dse to uncover new code by feeding \dse with the simplified trace: in~\cite{DebraySP15} this symbiosis turned out effective in cases that \dse alone could not handle. \puh{\tds} succeeded on code protected \puh{by state-of-the-art} VM obfuscators, as well as on four hand-written ROP programs.

We consider \cse, \dse, and \tds as they represent powerful tools available to adversaries, and embody concepts seen also in attacks against specific obfuscations (e.g.,~\cite{VMHUNT-CCS18}). Prior literature~\cite{banescu-acsac16,Banescu-USEC17,bardin-acsac19} uses \cse and \dse to evaluate and compare obfuscation techniques on goals akin to \goalone and \goaltwo, as both approaches are powerful and driven only by the semantics of the code \puh{(i.e., syntactic changes have little effect)}.
%: i.e., syntactic transformations have little effect, {\em including writing ROP code in standard ways} (\Cref{se:strengthening}).

%% TODO X-Force here??? later with ROPMEMU?

\subsubsection{ROP-Aware Techniques}
Expressing programs as ROP payloads affects analysis techniques that account for the syntactic representation of code. For instance, even commercial disassemblers and decompilers are not equipped to deal with this exotic representation and would fail to produce meaningful outputs for ROP chains~\puh{\cite{ROPOB-SecureComm17}}. Currently researchers have come up with two solutions to handle complex ROP code.

ROPMEMU~\cite{ropmemu} attempts dynamic multi-path exploration by looking for sequences that leak condition flags from the CPU status register, as they may take part in branching sequences (\Cref{ss:rop}): it flips their value and tries to generate alternate execution traces that explore new code. ROPMEMU is not the sole embodiment of this technique, \puh{seen in, e.g., crash-free binary exploration~\cite{xforce-usenix} and malware unpacking~\cite{rambo-dimva} research for RIP-driven code. ROPMEMU eventually} removes the ROP dispatching logic (i.e., the {\tt ret} sequences) and performs further simplifications, reconstructing a CFG representation. % new code \footnote{}

ROPDissector~\cite{ropscozzo} addresses shortcoming of ROPMEMU in branch identification, with a data-flow analysis for identifying sequences that build variable RSP offsets, so to flip all and only the operations taking part in the process. ROPDissector builds a ROP CFG highlighting branching points and basic blocks in a chain, and \puh{operates as} a static technique as it does not require a valid execution context as starting point. %can work also as

In our evaluation we will consider a combination of the two approaches, speculating on extensions tailored to our design.

%% file: design.tex
% !TEX root = ms.tex

\section{Program Encoding with ROP}
\label{se:encoding}

\begin{figure}[t!]
\centering
\includegraphics[width=0.73\linewidth]{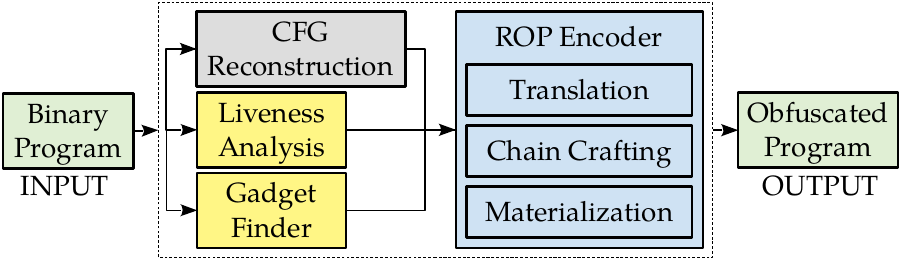}%775
\vspace{-1mm}
\caption{Architecture of the ROP rewriter.\label{fig:architecture}}
\vspace{-1.25mm}
\end{figure}

We design a binary rewriter for protecting compiled programs: the user specifies one or more functions of interest that the rewriter encodes as self-contained ROP chains stored in a data section of the binary. Our implementation supports compiler-generated, possibly stripped x64 Linux binaries. To ensure compatibility between ROP chains and non-ROP code modules, we intercept and preserve stack manipulations and use a separate stack for the chain.
This section details the design of the rewriter (\Cref{fig:architecture}), how it encodes generic functions as self-contained chains, and its present limitations.

\subsection{Geometry of a ROP Encoder}
\label{ss:geometry}

\subsubsection{Gadget Sources}
%\label{sss:gadget-sources}
The first decision to face in the design of a ROP encoder is where to find gadgets. These may be found in statically and dynamically linked libraries, in program parts left unobfuscated, or in custom code added to the program. We ruled out static libraries as a binary might not have any, and dynamic ones to avoid dependencies on specific library versions that must be present \puh{in any target system}.

%As for program code, code sizes as small as 20-100KB may already contain minimal gadget sets for exploitation~\cite{exploit-hardending}.
\puh{Exploitation research suggests that program code as small as 20-100KB may already contain minimal gadget sets for attacks~\cite{exploit-hardending}.}
% special
Our scenario however is \puh{ideal}: the possibility of controlling and altering the binary grants us more wiggle room compared to attack scenarios, as we can add missing gadgets---and most importantly create diversified alternatives---as dead code in the {\tt .text} section of the program. We thus pick gadgets from a pool of artificial gadgets combined with gadgets already available in program parts left unobfuscated.

\subsubsection{Rewriting} The second decision concerns deploying the encoder as a binary rewriter (as we do) or a compiler pass. Binary rewriting  can handle a larger pool of programs, including proprietary software and programs with a custom compilation toolchain, and builds on analyses that extract facts necessary to assist the rewriting. A compiler pass has some such analyses (e.g., liveness) already available during compilation, and possibly more control over code shape. However, in order to be able to rewrite an {\em entire} function, we believe a pass may have to operate as last step (modifying or directly emitting machine instructions) and/or constrain or rewrite several pieces of upper passes (e.g., instruction selection, register allocation). This would lead to a pass that is platform-dependent and that faces similar challenges to a rewriter while being less general.

\subsubsection{Control Transfers and Stack Layout}
\looseness=-1
Obfuscated functions get expressed in ROP, but may need to interact with surrounding components, calling (or being called by) non-ROP program/library functions or other ROP functions. In this respect native code makes assumptions on the stack layout of the functions, e.g., when writing return addresses or referencing stack objects in the scope of a function and its callees.

\begin{figure}[t!]
\centering
\includegraphics[width=0.715\linewidth]{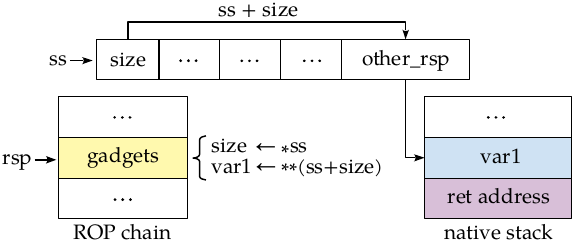}
\vspace{-1mm}
\caption{Reading a stack variable from top of native stack.\label{fig:rop-read-from-stack}}
\vspace{-1mm}
\vspace{-2.5mm}
\end{figure}

Reassembleable disassembling literature~\cite{uroboros,retrowrite,egalito,multiverse} describes known hurdles when trying to turn hard-coded stack references into symbols that can be moved around. In our design we instead preserve the original stack behavior of the program: we place the chain in a separate region, and rewrite RSP dereferences and value updates to use a {\tt other\_rsp} value that mimics how the original code would see RSP (\Cref{fig:rop-read-from-stack}).

This choice ensures a great deal of compatibility, and avoids that calls to native functions may overwrite parts of the ROP chain when executing. We keep {\tt other\_rsp} in a {\em stack-switching array} {\tt ss} that ensures smooth transitions between the ROP and native domains and supports multiple concurrently active calls to ROP functions, including (mutual) recursion and interleavings with native calls.

We store the number of active ROP function instances in the first cell of the array, making the last one accessible as {\tt *(ss+*ss)}. When upon a call we need to switch to the native domain, we use {\tt other\_rsp} to store the resumption point for the ROP call site, and move its old value in RSP so to switch stacks. Upon function return, a special gadget switches RSP and {\tt other\_rsp} again (\Cref{fig:rop-call}).

\subsubsection{Chain Embedding} Upon generation of a ROP chain, we replace the original function \puh{body} in the program with a stub that switches the stack and activates the chain. We opt for chains without destructive side effects, avoiding \puh{to have} to restore fresh copies across subsequent invocations. We place the generated chains at the end of the executable's \puh{\tt .data} section or in a dedicated one.

%\vspace{-1pt}
\subsection{Translation, Chain Crafting, and Materialization}
\label{ss:rop-rewriter-pipeline}
This section describes the rewriting pipeline we use in the ROP encoder of \Cref{fig:architecture}. Although we operate on compiled code, the pipeline mirrors typical steps of compiler architectures~\cite{compilers-book}: we use a number of support analyses (yellow and grey boxes) and {\em translate} the original instructions to a simple custom representation made of {\em roplets}, which we process in the {\em chain crafting} stage by selecting suitable gadgets for their lowering and then allocating registers and other operands. A final {\em materialization} step instantiates symbolic offsets in the chain and embeds the \scz{output} raw bytes in the binary.% as outlined at the end of the previous section. 

%\vspace{-1pt}
\subsubsection{Translation}
\label{sss:translation}
The unit of \puh{transformation} is the function. We identify code blocks and branches in it using off-the-shelf disassemblers (\textit{CFG reconstruction} element of \Cref{fig:architecture}): Ghidra~\cite{ghidra}  \puh{worked flawlessly} in our tests when analyzing indirect branches, and we support angr~\cite{angr} and radare2~\cite{radare2} as alternatives. We then translate one basic block at a time, turning its instructions into a sequence of roplets.

A {\em roplet} is a basic operation of one of the following kinds:
\begin{itemize}
\ifonlinereport
    \item {\em intra-procedural transfer}, for direct branches and for indirect branches coming from switch tables (\Cref{apx:a});
\else
    \item {\em intra-procedural transfer}, for direct branches and for indirect branches coming from switch tables (see~\cite{our-extended});
\fi
    \item {\em inter-procedural transfer}, for calls to non-ROP and ROP functions (including {\tt jmp}-optimized tail recursion cases);
    \item {\em epilogue}, for handling instructions like {\tt ret} and {\tt leave};
    \item {\em direct stack access}, when dereferencing and updating RSP with dedicated read or write primitives (e.g., {\tt push}, {\tt pop});
    \item {\em stack pointer reference}, when the original program reads the RSP value as source or destination operand in an instruction, or alters it by, e.g., adding a quantity to it;
    \item {\em instruction pointer reference}, to handle RIP-relative addressing typical of accesses to global storage in {\tt .data}; % the data segment; %used for accessing
    \item {\em data movement}, for {\tt mov}-like data transfers that do not fall in any of the three cases above;
    \item {\em ALU}, for arithmetic and logic operations.
\end{itemize}

One roplet is usually sufficient to describe the majority of program instructions. In some cases we break them down in multiple operations: for instance, for a {\tt mov qword [rsp+8], rax} we generate a stack pointer reference and a data movement roplet. To ease the later register allocation step, we annotate each roplet with the list of live registers\footnotemark{} found for the original instruction via \textit{liveness analysis}.

\footnotetext{A backward analysis deems a register {\em live} if the function may later read it before writing to it, ending, or \scz{making a call that may clobber it}~\cite{wcre02,delia-pldi18}.}

At this stage we parametrically rewire every stack-related operation to use {\tt other\_rsp}, and transform RIP-relative addressing instances in absolute references to global storage. % instances

%\vspace{-1pt}
\subsubsection{Chain Crafting}
\label{sss:crafting}
When the representation enters the chain crafting stage, we lower the roplets in each basic block by drawing from suitable gadgets for each roplet type (using the \textit{gadget finder} element of \Cref{fig:architecture}). For instance, to translate a conditional (left) or unconditional (right) intra-procedural transfer we combine gadgets to achieve:

% TODO fix this crappy layout
\begin{center}
\begin{footnotesize}
\begin{tabular}{>{\ttfamily}l || >{\ttfamily}l}
pop \{reg1\} \phantom{x}\#\# L & \\
mov \{reg2\}, 0x0 & \\
cmov\{ncc\} \{reg1\}, \{reg2\} \hspace{2pt} & \hspace{2pt} pop \{reg1\} \phantom{x}\#\# L \\
add rsp, \{reg1\} & \hspace{2pt} add rsp, \{reg1\} \\
\end{tabular}
\end{footnotesize}
\end{center}

where a gadget may cover one or more consecutive lines (so we omit {\tt ret} above). In both \puh{codes} the {\tt pop} gadget will read from the stack \scz{an operand} {\tt L} (placed as an immediate between the addresses of the first and second gadget) that represents the offset of the destination block. {\tt L} is a symbol that we materialize once the layout of the chain is finalized, similarly to what a compiler assembler does with labels.

Following the analogy, when choosing gadgets for roplets we operate as \puh{when in} the {\em instruction selection} stage of a compiler~\cite{compilers-book}, with {\tt \{regX\}} representing a virtual register, roplets the middle-level representation, and gadgets the \puh{low}-level one. When it comes to {\em instruction scheduling}, we follow the \puh{order} of the original instructions in the block. % same order

\begin{figure}[t!]
\centering
\includegraphics[width=0.765\linewidth]{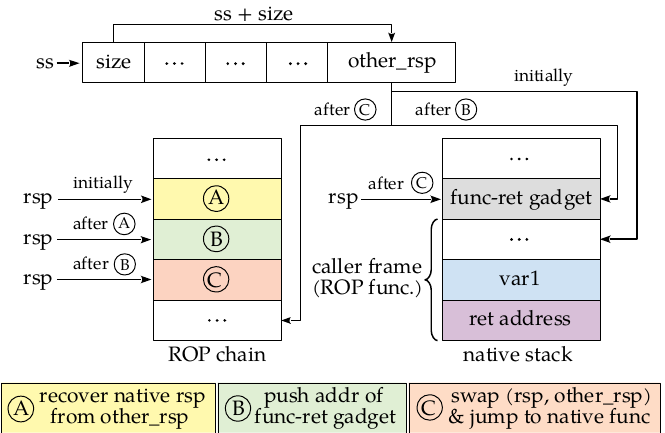}
\vspace{-0.6mm}
\caption{Call to a native function from ROP code.\label{fig:rop-call}}
\vspace{-1.5mm}
\end{figure}

Native function calls see a special treatment, as we have to switch stacks and set up the return address in a way to make another switch and resume the chain (\Cref{ss:geometry}). For the call we combine gadgets as \puh{in the following}:

\begin{verbatim}
  pop {reg1}  ## ss
  add {reg1}, qword ptr [{reg1}]  ## step A ends
  sub qword ptr [{reg1}], 0x8
  mov {reg2}, qword ptr [{reg1}]
  pop {reg3}  ## addr. of return gadget
  mov qword ptr [{reg2}], {reg3}  ## step B ends
  pop {reg2}  ## function address
  xchg rsp, qword ptr [{reg1}]; jmp {reg2} ## step C
\end{verbatim}

where we pop from the stack the addresses of: the stack-switching array, a {\em function-return gadget}, and the function to call. Gadgets may cover one or more consecutive lines, except for the last one which describes an independent single JOP gadget (\Cref{ss:rop}): {\tt xchg} and {\tt jmp} switch stacks and jump into the native function at once. \Cref{fig:rop-call} shows the effects of the three main steps carried by the sequence.

The called native function sees as return address (top \puh{entry} of its stack frame) the address of the function-return gadget. This is a synthetic gadget with a statically hard-wired {\tt ss} address that reads the RSP value saved by the {\tt xchg} at call time and swaps stacks again:

%\noindent
\vspace{-0.2em}
\begin{verbatim}
    mov {reg1}, ss; add {reg1}, qword ptr [{reg1}];
    xchg rsp, qword ptr [{reg1}]; ret
\end{verbatim}
\vspace{-0.2em}

For space limitations we omit details on the lowering of other roplet types: their handling \scz{becomes ordinary once} we translated RSP and RIP-related manipulations (\Cref{sss:translation}). 

\smallskip
{\em Register allocation} is the next main step: we choose among candidates available for a desired gadget operation by taking into account the registers they operate on and those originally used in the program, trying to preserve the original choices whenever possible. When we find conflicts that \puh{may clobber a register}, we use scratch registers when available (i.e., non-live ones) or \puh{spill it} to an inlined 8-byte chain slot as a fallback. We then ensure a {\em reconciliation of register bindings}~\cite{pin} at the granularity of basic blocks: when execution leaves a block, the CPU register contents reflect the expected locations for program values that are live in the remainder of the function. 

Another relevant detail is to preserve the status register if the program may read it later. While most instructions alter CPU flags, our liveness analysis points out the sole statements that may concur to a later read: whenever in between we introduce gadgets that pollute the flags\ifonlinereport\footnotemark{}\else\ifdefined\submittedpaper\footnotemark{}\fi\fi, we spill and later restore them.

\subsubsection{Materialization}
\label{sss:materialization}
At the end of the crafting stage a chain is almost readily executable. As its branching labels are still symbolic, we \puh{may} optionally rearrange basic blocks: then once we fix the layout the labels become concrete RSP-relative displacements. We then embed the chain in the binary, allocating space for it in a data section and replacing the original function code with a pivoting sequence to the ROP chain. The sequence extends the stack-switching array and saves the native RSP value, then the chain upon termination executes a symmetric unpivoting scheme (details in \apxref).

%\vspace{-2pt}
\subsection{Discussion}
\label{ss:design-discussion}
% on the capabilities of
Our design makes limited assumptions on the input code: it hinges on off-the-shelf binary analyses to identify intra-procedural branch targets, and obliviously translates stack accesses and dereferences to preserve execution correctness when interacting with the surrounding software stack.

Our implementation could rewrite a large deal of real-world programs (\Cref{sss:eval-coverage}), even when we supplied it code already protected by the control-flow flattening and/or (nested) VM obfuscations of the Tigress framework~\cite{tigress}. We experimentally observed (\Cref{sss:eval-coverage}) that the analyses of Ghidra are \puh{remarkably} effective in recovering intra-procedural indirect branch targets, which in several high-level languages derive from optimized switch constructs. Whenever those may fail, \puh{one} could couple the rewriter with a dynamic tracer for recovering the intended targets by running the original program using expected inputs. Transfers to other functions via indirect calls or tail jumps are instead straightforward, as the chain transfers control to the prologue of the callee as it happens with direct calls. 

% FOOTNOTE
\ifonlinereport
\footnotetext{This happens mainly when an intervening instruction involves RSP and we emit gadgets to do pointer arithmetic for stack pointer reference roplets.}
\else
\ifdefined\submittedpaper
\footnotetext{\cready{}{This happens mainly when an intervening instruction involves RSP and we emit gadgets to do pointer arithmetic for stack pointer reference roplets.}}
\fi
\fi

A limitation of the design, shared with static rewriting and instrumentation schemes~\cite{sok-dbi,retrowrite}, is lastly the inability to handle self-modifying and dynamically generated code.

As for register conflicts, the high number of x64 registers give us wiggle room to perform register renaming within blocks with modest spilling. An area larger than the 1-word one we use may help with code with very high register pressure cases (\Cref{ss:obfuscation-eval-perf}) or 32-bit implementations; instruction reordering and function-wide register renaming may also help.

The implementation incurs two main limitations that one can address with moderate effort. The spilling slots and the {\tt ss} array area are not thread-private, but we \puh{may} recur to thread-local storage primitives. Rewritten binaries are compatible with address space layout randomization for libraries, while the body of the program is currently loaded at fixed addresses. To ship position-independent executables we may add relocation information to headers so to have the loader patch gadget addresses in the chains, or use the online patching for chains from~\cite{ropneedle} to have the program itself do the update. % a covert?

In terms of compatibility with ROP defenses of modern operating systems, our context is different to an exploitation one where the program stack gets altered and the choice of gadgets is limited. On Windows, for instance, our stack switching upon API calls would already comply with the RSP range checks of \textit{StackPivot}~\cite{N-SISY15}; our liberty to synthesize gadgets would be decisive against \textit{CallerCheck}, which checks if the instruction preceding an API's return address is a call~\cite{N-SISY15}. We refer to prior work~\cite{ropneedle} for details. A potential issue, which may require the user to whitelist the program, could be instead coarse-grained defenses that monitor branches~\cite{stitching} or micro-architectural effects~\cite{ml-rop}. However, those are yet to become mainstream as they face robustness and accuracy issues.

Finally, our readers may question if the use of ROP introduces obvious security risks. An attacker needs a write primitive pointing to a chain in order to alter it. In our protected programs, ROP-encoded parts use write operations only for spilling slots, and those cannot go out of bounds. Non-ROP parts never reference chains in write (or read) operations: an attacker would thus have to search for an arbitrary memory write primitive in such parts. Its presence, however, would be an important source of concern even for the original program. Our implementation also supports the generation of read-only chains, which use a slightly longer spilling machinery.

%% file: strengthening-rop.tex
% !TEX root = main.tex
\section{Strengthening ROP Programs}
\label{se:strengthening}

%\smallskip % COSA CI FACEVI QUI A TOGLIERCI SPAZIO!!! :-)))
In the furrow of prior works (e.g.,~\cite{DebraySP15, ropmemu, synthia}) that highlighted the hindrances from the ROP paradigm to reverse engineering attempts, one could anticipate that the design of \textsection\ref{se:encoding} may challenge manual deobfuscation and code understanding attempts. The common thread of their observations is that the exoticism of the representation---ROP defines a {\em weird machine}~\cite{franz-sp-magazine}---disturbs humans when compared to native code. The \puh{rewriter} makes use of all motivating factors for ROP that we outlined in~\Cref{se:intro}, such as destructured control flow and diversity and reuse of gadgets \puh{(including gadget confusion that we describe next)}.

Quantifying the effectiveness of an obfuscation is however a difficult task, as it depends not only on the available tools, but also on the knowledge of the human operating them~\cite{banescu-acsac16}. A well-established practice in the deobfuscation literature is to measure the resilience to automated deobfuscation techniques, which in most attack scenarios are the fulcrum of reverse engineering attempts and ease subsequent manual inspections~\cite{SurveyCSUR16}. 

% : and the lack of specific analysis tools for the paradigm (although recent years have seen important advancements~\cite{gao3,ropmemu,ropscozzo}). 
%%Several works~\cite{puh,puh2,puh3,puh4} argued the difficulties that a human agent may incur when analyzing ROP code due to the exoticism of the representation (ROP defines a {\em weird machine}~\cite{franz-sp}) when compared to native code. %Current ROP analysis tools are also not conceived for {\em shielded} payloads.

%We think that measuring manual deobfuscation effort, in principle challenging to model and in turn to evaluate~\cite{xxx,yyy}, would however only tell a partial story.

%Following established practices~\cite{banescu16,bardin19}

\smallskip
\textbf{\em ROP encoding alone is not sufficient for obfuscation.} We find control transfers between basic blocks to be its weak link.

\puh{Even when diversifying the used gadget instances, an attacker aware of the design may follow the ROPMEMU approach (\Cref{ss:deobfuscation-solutions}) to spot in an execution trace what gadgets add variable quantities to RSP (thus exposing basic blocks),}untangle {\tt ret} instructions from the original control flow of the program, and assemble a dynamic CFG from multiple traces.

%Even when diversifying gadget instances for the encoding, an attacker aware of the design may try in the spirit of ROPMEMU (\Cref{ss:deobfuscation-solutions}) to look for gadgets that in an execution trace add a variable quantity to RSP as basic blocks delimiters, untangle {\tt ret} instructions from the original control flow of the program, and assemble a dynamic CFG from multiple traces. % one or more traces

Protecting control transfers is equally critical in the face of the most effective general-purpose semantics-aware techniques like \cse, \dse and \tds, which try to reason on the parts essential for program functionality while sifting out the irrelevant obfuscation constructs and instructions~\cite{banescu-acsac16,DebraySP15}, such as side effects and dynamically dead portions from gadgets.
%These benefits may vanish when adversary aware of the implementation details of the rewriter devises heuristics, in the spirit of~\cite{ropmemu,ropscozzo}, that look for possible jump points (i.e., gadgets that alter the value of RSP before executing  {\tt ret}) that naturally delimit basic blocks in the ROP chain, and untangle such blocks from the ROP gadget dispatching and immediate-operand loading logic.
%
%Protecting control transfers is critical also in the face of the most effective general-purpose deobfuscation techniques, which try to reason on the parts essential for program functionality while sifting out the irrelevant obfuscation constructs and instructions~\cite{banescu-acsac16,debray-sp15}.

% obvious
One way to hinder the automated approaches of \Cref{ss:deobfuscation-solutions} would be to target weaknesses of each technique individually. For instance, researchers proposed hard-to-solve predicates for \cse (e.g. MBA expressions~\cite{biondi17}, cryptographic functions~\cite{sharif-CHECK}), and code transformations that impact concolic variants like \dse too~\cite{debray-ccs15}. But an experienced attacker can symbiotically combine methods to defeat this approach, for instance using TDS or similar techniques (e.g., program synthesis for MBA predicates~\cite{biondi17}) to feed DSE with tractable traces as in~\cite{DebraySP15}. % TODO MBA predicates or expressions?

In this section \puh{instead} we present three rewrite predicates, naturally meshed with RSP update actions, that bring protection against generic, increasingly powerful automated attacks that cover the principled classes \puh{\surfacesome{1-2-3}} from \Cref{ss:principles}. We then introduce gadget confusion and share some general reflections.
%  \surfaceone, \surfacetwo, and \surfacethree

% PEZZI POSSIBILMENTE UTILI
% - For each predicate we evaluate its resistance against the state-of-the-art deobfuscation approaches mentioned before.
% \footnotetext[6]{\scz{Identified with a backward taint analysis akin to dynamic program slicing~\cite{DebraySP15}.}}

\subsection{Predicate {\small\predone}: Anti-ROP-Disassembly}
\label{sss:archer}

\begingroup
\setlength{\thickmuskip}{2.5mu}
Our first predicate uses an array of opaque values~\cite{CollbergBook} to hide branch targets (\surfaceone). The array contains seemingly random values generated such that a periodic invariant holds, and backs the extraction of a quantity $a$ that we use to compute the displacement in the chain for one of the $n$ branches in the code. Suppose we need to extract $a$ for branch $b\in\{0..n-1\}$: starting with cell $b$, in every $p$-th cell of the array we store a random number $q$ such that $q\equiv a \,\textrm{mod}\,m$, with $m>n$ and $p$ chosen at obfuscation time.
\endgroup

%\vspace{-1pt}
\definecolor{gr1}{gray}{0.85}
\definecolor{gr2}{gray}{0.65}
\definecolor{gr3}{gray}{1.00}
%\vspace{-0.15em}
\begin{center}
\adjustbox{max width=0.98\columnwidth}{
\newcolumntype{x}{>{\columncolor{gr1}}c}
\newcolumntype{y}{>{\columncolor{gr2}}c}
\newcolumntype{z}{>{\columncolor{gr3}}c}
\begin{tabular}{|x|y|z|x|y|z|x|y|z|x|y|z|}
\hline
10 & 19 & 34 & 45 & 54 & 62 & 66 & 33 & 6 & 59 & 61 & 20 \\
\hline
\end{tabular}
}
\end{center}
%\vspace{-0.15em}

\newcommand{\scozbox}[1]{\fcolorbox{black}{#1}{\rule{0pt}{1pt}\rule{1pt}{0pt}}}
\begingroup
\setlength{\thickmuskip}{2.5mu}
%In the example a
\noindent
Above we encoded information for $n\mathord{=}3$ branches  using $p\mathord{=}4$ repetitions and $m\mathord{=}7$. For the branch with ordinal $1$ we wanted to memorize $a=5$: every cell colored in dark gray thus contains a value $v$ such that $v\,\textrm{mod}\,7$ \puh{equals} $5$. % always equals

During obfuscation we use a period of size $s>n$, with a fraction of the cells containing garbage. We also share a valid cell among multiple branches, so to avoid encoding unique offsets that may aid reversing. To this end we divide an RSP branch offset $\delta$ in a fixed part $a$ encoded in the array and a branch-specific part $\delta-a$ computed by the chain, then we compose them upon branching.

% purposes
% can bring
This implies that for static disassembly  an attacker should recover the array representation and mimic the computations made in every chain segment to extract $a$ and compute the branch-specific part. While this is possible for a semantically rich static technique like \cse, periodicity comes to the rescue as it \puh{brings} {\em aliasing}: every $p$-th cell is suitable for extracting $a$. Our array dereferencing scheme takes the form of:% an expression:

%\vspace{-0.2em}
\begin{center}
$a = A[f(x)*s+n]\,\textit{mod}\,m$
\end{center}
%\vspace{-0.2em}

where$\,f(x)\,$depends on the program state and returns a value between $0$ and $p-1$. Its implementation opaquely combines the contents of up to 4 registers that hold input-derived values. SE will thus explore alternative input configurations that ultimately lead to the same $\textrm{rsp}\mathrel{+}=\delta$ update; reducing their number by constraining the input would lead instead to missing later portions of program state.

% is capable of implementing
Whenever an attacker may attempt a {\em points-to} analysis~\cite{Smaragdakis-pointer-analysis-tutorial} over $\textrm{rsp}\mathrel{+}=\delta$, we believe a different index expression based on user-supplied or statically extracted facts on input value ranges would suffice to complicate such analysis significantly. % be sufficient,  such an
%Incidentally, this introduces an artificial implicit control dependency~\cite{debray-sp15} on the input (C).
\endgroup

\subsection{Predicate {\small\predtwo}: Preventing~Brute-Force Search}
\label{sss:fisto}
Our second predicate introduces artificial data dependencies on the control flow, hindering dynamic approaches for brute-force path exploration (\surfacetwo) that flip branches from an execution trace.  While these techniques do not help in secret finding (\goalone) as they neglect data constraints \puh{(\Cref{ss:principles})}, they may be effective when the focus is code coverage (\goaltwo).
%These techniques do not help in secret finding scenarios as they explore code under violated data constraints, but are powerful when the focus is code coverage.
%
%Predicate

\predone is not sufficient against \surfacetwo: an attacker can record a trace that takes a conditional branch shielded by \predone, analyze it to locate the flags set by the instruction that steered the program along the branch, flip them, and reveal the other path~\cite{ropscozzo}.

%if an attacker records a trace taking a conditional branch shielded by \predone, they may analyze the trace to locate the flags set by the instruction that steered the program along the branch, flip them, and reveal the other path~\cite{ropscozzo}.

Without loss of generality, let us assume that a $\textit{cmp a, b}$ instruction determines whether the original program should jump to location {\tt L} when $a==b$ and fall through otherwise. We introduce a data dependency that breaks the control flow when brute-force attempts leave its operands untouched. 
% $\alpha\cdot x$. 
As we translate the branch in ROP, in the block starting at {\tt L} we manipulate RSP with, e.g., $\textit{rsp} \mathrel{+}= x*(a-b)$, so that when brute-forcing it without changing the operands, $(a-b)\mathrel{!}=0$ and RSP flows into unintended code by \puh{some offset multiple of $x$}. Similarly, on the fall-through path we manipulate RSP with, e.g., $\textit{rsp} \mathrel{+}= x*(1-\textit{notZero}(a-b))$, where $\textit{notZero}$ is a flag-independent computation\footnotemark{} so the attacker cannot flip it.

% thought of (1 rigo extra)
Different formulations of opaque updates are possible. Whenever an attacker may attempt to learn and override \puh{updates} locally, we \scz{figured} a future, more covert \predtwo variant that \puh{encodes offsets for branches using opaque expressions based on value invariants (obtainable via value set analysis~\cite{value-set-analysis}) for some variable that is defined in an unrelated CFG block}. 

%, using value range information~\cite{value-set-analysis} for a visible variable defined in an unrelated CFG block, encodes offsets for later branches using such value opaquely.
%where such value is still visible.

%As future extension we  thought of a non-local, more covert P2 variant using value range information~\cite{value-set-analysis} for a variable from an unrelated CFG block to shield later branches where its value is still visible.% (and without intervening redefinitions).

%One way to make this predicate stronger could be to infer and use invariants on the program state using static analyses. For instance, if we know the value range for a variable at a given location in the control flow graph, in the absence of redefinitions we can use this information to shield the branching points dominated by it.

\subsection{Predicate {\small\predthree}: State Space Widening}
\label{sss:forvizsec}
Our third predicate brings a path-oriented protection that artificially extends the program space to explore and is coupled with data (and \puh{optionally} control) flows of the program, so that techniques like \tds (\surfacethree) cannot remove it without knowledge of the obfuscation-time choices. \predthree comes in two variants.

%$k$ loops to recompute a function parameter or some input-derived program value, introducing $2^{n\cdot k}$, where $n$ is the size in bits of the value.

%, which generalizes and boosts the {\em range divider} idea from~\cite{banescu}
% recently proposed in

% char fc
% uses in the code
The first variant is an adaptation of the {\sc For} predicate from~\cite{bardin-acsac19}. The idea is to introduce state forking points using loops, indexed by input bytes, that opaquely recompute available values that the program may use later. In its simplest formulation, {\sc For} replaces occurrences of an input value {\tt char c} with uses of a new {\tt char fc} instantiated by {\tt for (i=0; i<c; ++i) fc++}. Such loop introduces $2^8$ artificial states to explore due to the uncertainty on the value of {\tt c}.

\puh{The work explains that} targeting 1-byte input portions brings only a slight performance overhead, and choosing independent variables for multiple {\sc For} instances optimizes composition for state explosion. %To keep it time-efficient its use is better restricted to 1-byte input portions, then targeting independent variables when composing {\sc For} instances optimizes state explosion.
It also argues how to make {\sc For} sequences resilient to pattern attacks, and presents a theorem for robustness against taint analysis and backward slicing, considered for forward and backward code simplification \puh{attacks}, respectively (the TDS technique we \puh{use} has provisions for both~\cite{DebraySP15}).

While we refer to it for the formal analysis, for our goals suffice it to say that when the obfuscated variable is input-dependent (for tainting) and is related to the output (for slicing), such analyses cannot simplify away the transformation.

% TODO: slicing di angr?
During the rewriting we use a data-flow analysis to identify which live registers contain input-derived data ({\em symbolic} registers) and \scz{may} later concur to program outputs\footnotemark{}. We then introduce value-preserving opaque computations like in \puh{the} examples below (the right one is adapted from~\cite{bardin-acsac19}): % (with the one on the right borrowed and adapted from~\cite{bardin-acsac19}):

%\vspace{-1pt}
\begin{center}
\adjustbox{max width=0.95\columnwidth}{%PUUUHHH
\begin{tabular}{>{\ttfamily}l | >{\ttfamily}l}
\multirow{6}*{\shortstack[l]{%
// clear last byte\\%
dead\_reg \&= 0xAB00;\\%
for (i=0; i<(char)sym; ++i)\\%
\hphantom{xx}dead\_reg++;\\%
sym |= (char)dead\_reg;%
}} & \multirow{6}*{\shortstack[l]{%
dead\_reg = 0;\\%
for (i=0; i<(char)sym; ++i)\\%
\hphantom{xx} if (i\%2) dead\_reg--;\\%
\hphantom{xx} else dead\_reg+=3;\\%
if (i\%2) dead\_reg-=2;\\%
sym = (sym\&0xF..F00)+dead\_reg;%
}} \\
\\
\\
\\
\\
\\
\end{tabular}
}
\end{center}
%\vspace{-1pt}

% TODO questo claim va verificato molto bene, tuttavia promising
These patterns significantly slow down \cse and \dse engines, but also challenge approaches that feed tractable simplified traces to \dse. While one may think of detecting and propagating constant values in the trace, the \tds paper~\cite{DebraySP15} explains that doing it indiscriminately may oversimplify the program: in our scenario it may remove {\sc For} but also pieces of the logic of the original program elsewhere. To avoid oversimplification the \puh{\tds} authors restrict constant propagation across input-tainted conditional jumps, which is exactly the case with {\tt dead\_reg} and {\tt sym} in the examples above.  
%~\cite{debray-sp15}

%  in~\cite{DebraySP15}
The \puh{authors} suggest, \puh{as} a general way to hamper semantics-based deobfuscation approaches like \tds, \puh{to deeply entwine} the obfuscation code with the original input-to-output computations. \puh{They also state that at} the time obfuscation tools had not explored this avenue, possibly for the difficulties in preserving observable program behavior~\cite{DebraySP15}. 

%\smallskip
\begingroup
\setlength{\thickmuskip}{2.5mu}
%We propose a new P3 variant that
%% taken later
Our second \predthree variant is new and moves in this direction. Instead of recomputing input-derived variables, we use them to perform {\em opaque updates} to the array used by \predone. Updates include adding/subtracting quantities multiple of $m$, swapping the contents of two related cells from different periods, or combining the contents of two cells $i$ and $j$ where $a\equiv A[i] \,\textrm{mod}\,m$ and $b\equiv A[j] \,\textrm{mod}\,m$ to update a cell $l$ where $(a+b)\equiv A[l] \,\textrm{mod}\,m$. For \dse\puh{-alike} path exploration approaches the effect is tantamount to the {\sc For} transformation described above. For trace simplification it introduces implicit flows, with {\em \puh{fake control} dependencies} between program inputs and branch decisions \puh{taken later} in the code: \tds cannot simplify them without explicit knowledge of the invariants. %semantics-based trace deobfuscation
\endgroup

% FOOTNOTES
\ifonlinereport
\footnotetext[3]{Example: $\textit{notZero}(n)\mathrel{:}={\sim}({\sim }n\,\&\,(n+{\sim}0)){\gg}31$ for 32-bit data types.} 
\else
\ifdefined\submittedpaper
\footnotetext[3]{Example: $\textit{notZero}(n)\mathrel{:}={\sim}({\sim }n\,\&\,(n+{\sim}0)){\gg}31$ for 32-bit data types.} 
\else
\footnotetext[2]{Example: $\textit{notZero}(n)\mathrel{:}={\sim}({\sim }n\,\&\,(n+{\sim}0)){\gg}31$ for 32-bit data types.} 
\fi
\fi
%\footnotetext{Example: $\textit{notZero}(n)\mathrel{:}={\sim}({\sim }n\,\&\,(n+{\sim}0)){\gg}31$ for 32-bit data types.} % While other branchless sequences are possible, ROP also eases diversity when instantiating them.
\footnotetext{To this end we use the symbolic execution capabilities of angr~\cite{angr}.}

\subsection{Gadget Confusion}
\label{se:gadget-confusion}
% meaning that we can use gadgets whose instructions concur to implementing a predicate or have no effect depending on the surrounding chain portion
ROP encoding brings several advantages when implementing \predsome{1-2-3}. Firstly, it offers significant leeway for diversifying the gadget instances we use to instantiate them. We combine this diversity with {\em dynamically dead} instructions\puh{: we can use gadgets whose each instruction either concurs to implementing a predicate or has no effect depending} on the surrounding chain portion. This helps in instantiating many variants of a pattern, challenging syntactic attacks aware of the design.

\looseness=-1
However, a unique advantage of ROP, as we observed in \Cref{se:intro}, is the level of indirection that it brings: this complicates pattern attacks that look for specific instruction bytes, since code is not in plain sight, and attackers need to extract the instruction sequences as if executing the program. What they see are bytes belonging to either gadget addresses or data operands. They may, however, attempt analyses that look for byte sequences resembling addresses from code regions (i.e., plausible gadgets) and try to speculatively execute the chain from there~\cite{Michalis-MALWARE11,ropscozzo}. By trying it at every plausible point, this may eventually reveal some chain portions, nonetheless short thanks to \predsome{1-2}.

% where
%  expected value
This is \puh{when} {\em gadget confusion} enters the picture. Firstly, we can transform data operands in the chain to look like gadget addresses, having then gadgets recover the \puh{desired values} at run time (e.g., subtracting two addresses to obtain a constant, applying bitmasks, shifting bits, etc.). This is possible as we control both the layout of the binary (for the addresses) and the pool of artificial gadgets (for the manipulations). Now that virtually every 8-byte chain stride looks like a gadget address, we introduce unaligned RSP updates at random program points, adding a quantity $\eta$ s.t. $\eta\,\textrm{mod}\,8\mathrel{!}=0$. In the end, the attacker may have to execute speculatively at every possible chain offset, obtaining instructions that may or may not be part of the intended execution sequence. We believe such gadget confusion makes pattern attacks on \puh{our} chains even harder.

\subsection{Further Remarks}
\label{ss:discussion-predicates}
The instantiation of \predsome{1-2-3} is naturally entwined with RSP dispatching: directly for \predsome{1-2}, and indirectly for \predthree through array updates. In the rewriter, \predone replaces the RSP update sequence we showed in \Cref{sss:crafting}, while \predtwo operates on the fall-through and target blocks of a branch. Finally, the rewriter can apply either \predthree variant to a user-defined fraction {$k$} of the original program points when lowering the associated roplets.
% TODO? Left out pieces
% - Finally, for \predsome{1-2} arithmetic and data movement operations become control flow transfers in disguise, granting more opportunities compared to classic means for RIP manipulation.

% TODO il problema di Bardin....

%\marco{ROP eases their instantiation in a number of ways.}{%
%While their instantiation is naturally entwined with RSP dispatching (directly for \predsome{1-2}; indirectly for \predthree through array updates), the very ROP encoding brings further advantages.} Adversaries cannot attempt static pattern attacks (code is not in plain sight, \Cref{se:intro}), but need to extract the instruction sequences as if executing the program. \marco{They also need}{In doing so they have} to discard \scz{{\em dynamically dead} instructions, as in the encoding we may use gadgets whose (non)-dead parts} change depending on the surrounding context. Combined with gadget diversity, such reuse helps in instantiating variants of a pattern. Finally, for \predsome{1-2} arithmetic and data movement operations become control flow transfers in disguise, granting more \scz{opportunities} compared to classic means for RIP manipulation.

% extensions
Each predicate targets a main attack surface, but positive externalities are also present. \predtwo can protect against possible linear/recursive disassembly algorithms for ROP (\surfaceone), but will not withstand \cse-based disassembly. In \Cref{ss:obfuscation-eval-perf} we discuss how \predone can slow down state exploration (\surfacethree) by indirectly putting pressure on the memory model of a \puh{\cse or \dse engine}.% symbolic executor.

\puh{Finally, with the second \predthree variant we used ROP control transfer dynamics to introduce also fake control dependencies.}
%Also, ROP control transfer dynamics offer opportunities for introducing opaque data dependencies, as we did with \predthree.

%% file: related.tex
% !TEX root = main.tex

\section{Other Related Works}
\label{se:other-related}

Prior research explored ROP for software protection goals orthogonal to obfuscation: tamper checking of selected code regions through chains that use gadgets from such regions~\cite{parallax}, covert watermark encoding~\cite{gao-asiaccs16}, and steganography of short code~\cite{ropsteg-codaspy14}. Each of them could complement our design, especially \cite{parallax} for \puh{checking} code integrity of non-obfuscated parts.

ROPOB~\cite{ROPOB-SecureComm17} is a lightweight obfuscation method to rewrite transfers between CFG basic blocks using ROP gadgets. It considers standard disassembly algorithms as adversary (a \puh{``lighter''} \surfaceone case), and does not withstand static attacks like \cse (\surfaceone) or ROPDissector (\surfacetwo), nor dynamic ones like \dse or \tds (\surfacethree). ROPOB leaves data manipulation instructions in plain sight, whose rewriting poses several challenges (\textsection\ref{ss:rop-rewriter-pipeline}).

VM deobfuscation attacks like Syntia~\cite{synthia} and VMHunt~\cite{VMHUNT-CCS18} intercept and simplify (\surfacethree) dispatching and opcode handling sequences. They do not apply directly to ROP chains, and embody flavors of the agnostic and general approach of \tds.

movfuscator~\cite{movobfuscator} is an extreme instance of the weird machine concept, rewriting programs using only the Turing-complete {\tt mov}~instruction. Kirsch et al. present~\cite{demov-paper} a \puh{custom} linear-sweep algorithm (\surfaceone) that recovers the CFG \puh{by} targeting logic dispatching elements \puh{used for} the very encoding.

%% file: evaluation.tex
% !TEX root = ms.tex

\section{Evaluation}
\label{se:evaluation}

We arrange our experimental analysis in three parts. We first study the efficacy of our techniques against prominent solutions for \surfacesome{1-2-3} (\Cref{ss:eval-deobfuscation}), confirming the theoretical expectations. We then study the resource usage of viable deobfuscation attacks using a methodology adopted in previous works~\cite{banescu-acsac16,bardin-acsac19}, and put such numbers into perspective with VM-obfuscated\footnotemark{} counterparts (\Cref{ss:obfuscation-eval-robustness}). Finally, we analyze the applicability of our \puh{method} to real-world code (\Cref{ss:obfuscation-eval-perf}).

\begin{table}[t!]
\begingroup
\setlength{\thickmuskip}{1.5mu}
\begin{small}
\begin{center}
\resizebox{0.433\textwidth}{!} {
\begin{tabular}{|l|l|}\hline
 {\sc Setting} & {\sc Description} \\ \hline\hline
%\ropother{x} & ROP obfuscation with \predthree \\
 %\hline
 \multirow{3}{*}{\ropobf{k}} & ROP obfuscation with \predthree inserted at a fraction of\\ %ROP obfuscation with predicates P1 and P2 enabled on the full program, while
& program points $k \in \{0, 0.05, 0.25, 0.50, 0.75, 1.00\}$  \\
& \puh{and with \predone instantiated with $n\mathord{=}4$, $s\mathord{=}n$, $p\mathord{=}32$} \\
\hline
 {\sc \normvm{n}} & $n$ layers of VM obfuscation with $n \in \{1, 2, 3\}$ \\
 \hline
 \multirow{2}{*}{\impvm{n}{x}} & $n$ layers of VM obfuscation with implicit flows used \\
  & for every VPC load at layer(s) $x \in$ \footnotesize{\{first, last, all\}} \\
%\hline
%{\sc {\sc $n$VM} + ROP$_{k}$} & Virtualization obfuscation as in {\sc $n$VM}, further\\
%& obfuscated using {\sc ROP$_{k}$} \\
 \hline
\end{tabular}
}
\end{center}
\end{small}
\vspace{-2.5mm}
\caption{Terminology for obfuscation configurations.\label{tab:variants}}
%\vspace{-7mm}
\endgroup
\end{table}

We ran the tests on a Debian 9.2 server with two Xeon E5-4610v2 and 256 GB of RAM.
\ifonlinereport
Appendix \Cref{apx:a} contains 
\else
Our online technical report~\cite{our-extended} details
\fi
the settings we used to generate our 72 test functions and the VM variants with Tigress, and more implementation details. The rewriter currently consists of \textasciitilde3K Python LOC. 

% they are
\Cref{tab:variants} details configuration naming for the main ROP and VM experiments. For the latter we try multiple layers of nested virtualization as \puh{this is} known to slow down \cse and \dse-based attacks~\cite{salawan,bardin-acsac19}, and use a Tigress predicate that adds implicit flows to  virtual program counter (VPC) loads: those frustrate taint analysis-based simplifications and also create many redundant states whenever VPC becomes symbolic.

% FOOTNOTE
\footnotetext{We do not consider commercial tools like VMProtect for two reasons: they offer little control over the transformations (but may rather combine many at once), and add tricks and bombs~\cite{DebraySP15} to break deobfuscation solutions by targeting implementation gaps instead of their methodological shortcomings.}

\subsection{Efficacy of ROP Strengthening Transformations}
\label{ss:eval-deobfuscation}

The techniques presented in \Cref{se:strengthening} should intuitively raise the bar to existing automated attacks, and hinder symbiotic combinations between them. We now study how each automated approach feels the effects of \puh{each technique individually} already on small program instances, discussing also \puh{design-aware} enhancements we tried for ROP tools. In the end, \dse emerges as the one and only viable option for our attacker.%: in \Cref{ss:obfuscation-eval-robustness} we will 
%\end{mybox}

We leverage the Tigress framework~\cite{tigress} to generate functions appropriate as reverse engineering targets with a desired complexity and structure. Tigress will also annotate CFG split and join points with probes to help us measure code coverage. %We test the effects of \predone, \predtwo, and \predthree individually.

\subsubsection{\bf\em General Attacks} In the context of general-purpose automated attacks, we consider angr~\cite{angr} as \cse engine, S2E~\cite{S2E} for \dse, and the \tds implementation released by its authors. 
\begingroup
\setlength{\thickmuskip}{2.5mu}
Let us start with \cse. For \predone we consider a function with control structure~\cite{tigress} \texttt{for (if (bb 4) (bb4))} having 4 mathematical computations per block, 15 loop iterations, and a single {\tt int} as input. In \puh{a} ``\ropother{1}'' version we encode in the array for \predone $n\mathord{=}4$ $\delta$-offsets\puh{,} with no garbage entries ($s=n$) and $p=32$ repetitions, for a total of $128$ cells populated statically. 

% TODO FIX ME HERE
To explore enough paths to hit all coverage points (\goaltwo), angr took a time in the order of seconds for the native function, and over 4500 seconds for \ropother{1}. % (tens of seconds for the \normvm{1} setting).
%The time angr took to explore sufficient paths to hit all coverage points (\goaltwo) was in the order of seconds for the original version, tens of seconds for a \normvm{1} setting, and over 4500 seconds for ROP-P1.
The aliasing \predone induces on RSP updates for branching slows angr down significantly already for little code, as the SMT solver sees increasingly complex expressions over RSP. Aliasing reverberates on secret finding (\goalone) too: with a simpler \texttt{for (for (bb 4))} code, angr cracked the secret in the order of seconds for the original code, and over 5 hours for \ropother{1}.
% (1VM: hundreds of seconds).
Other configurations of variable complexity confirmed these trends.
%These trends were confirmed in other configurations of variable \marco{complexity that we tried}{complexity}.
When we tested \predthree shielding a single program point per basic block, 24 hours were not sufficient for angr to crack the secret. These results suggest \cse may not be readily suitable against our approach.
\endgroup

As for \dse, in the experiments \predone impacted it slightly and only for \goaltwo: the reason is that S2E benefits from concrete input values when picking the next path to execute. For \predthree we obtained two confirmations: its two variants bring similar time increases, and while higher $k$ fractions of shielded program points inflate \puh{the state space possibly more}, code with small input space may not always offer sufficient independent sources (i.e., symbolic registers) for optimal composition of \predthree instances. We postpone a detailed analysis of \puh{the} induced overheads to~\Cref{ss:obfuscation-eval-robustness} \puh{as we consider larger code instances}.

\predone and \predthree resist \tds by design. The \puh{tested} output traces kept non-simplifiable (\Cref{sss:forvizsec}) implicit control dependencies from having a tainted input value determine a jump target: as those are pivotal to put pressure on \dse, combining \dse with \tds-simplified input traces~\cite{DebraySP15} would not ease attacks.

Summarizing, \predone and \predthree effectively raise the bar for \surfaceone and \surfacethree attacks, respectively: \cse and \tds look no longer useful already for little code. \predtwo and gadget confusion target syntactic approaches, unlike the semantics-aware attacks we considered above: we address them next in the ROP-aware domain.

\subsubsection{\bf\em ROP-Aware Attacks}
\label{sss:fistato}
To analyze ROP payloads we use and extend ROPDissector to start from a memory dump of the program taken when entering the chain of interest: in this configuration it operates as a hybrid static-dynamic analysis and surpasses ROPMEMU in branch analysis and flipping capabilities. With ROPDissector now embodying a full-fledged \puh{ROP-\surfacetwo}  approach, we test if it can help with \goaltwo, while \goalone is out of scope as \surfacetwo recovers code but neglects data constraints.

Backing our expectations, shielding branches with \predtwo in the rewriting makes ROPDissector fail in revealing any basic block other than those the input used for the test reveals.
We tried to further extend ROPDissector by using its gadget guessing technique (a ROP-educated form of pattern matching~\cite{ropscozzo}) to reveal new blocks by executing the chain at different start offsets. Our gadget confusion however makes such analysis explode, with many short and unaligned candidate blocks that are difficult to distinguish from \predtwo-protected true positives.

We conclude this \puh{part} by stressing the importance of \puh{conceiving all of} our protections. \predone impacts ROPDissector only if no dump is supplied, and \predthree does not affect it directly. Hence, without \predtwo an attacker could have used ROPDissector or a similar tool to aid semantic attacks in code coverage scenarios.

% (for the same reasons, \predone and \predthree have insignificant effects on the CFG blocks ROPDissector can find).

%In \edit{the preliminary}{this set of} experiments DSE was only slightly impacted by P1 for coverage as we experimented with S2E, which benefited from concrete input values when picking the next path to execute. Then from the P3 tests we obtained two confirmations: (i) its two variants bring similar time increases, and (ii) while higher $k$ fractions of shielded program points inflate state space widening, code with a small input space may not always offer sufficient independent (symbolic register) sources when composing P3 instances.

%For an attacker interested in secret finding (\goalone) or code coverage (\goaltwo) goals the three predicates from \Cref{ss:countering-attacks} represent hurdles for using available deobfuscation tools effectively. We test \marco{them}{the systems} individually using small program instances.

\subsection{Measuring Obfuscation Resilience} 
\label{ss:obfuscation-eval-robustness}

We now measure the amount of resources required to carry automated attacks for secret finding (\goalone) and code coverage (\goaltwo) over synthetic functions from an established methodology. We ask Tigress to generate 72 non-cryptographic hash functions with 6 control structures analogous to the most complex ones from an influential obfuscation work~\cite{Banescu-USEC17}, input sizes of $\{1, 2, 4, 8\}$ bytes, and three seeds (details in \apxref).

We exclude techniques that were ineffective on smaller inputs like \cse, and restrict our focus to \dse (recently~\cite{bardin-acsac19}  makes a similar choice). \dse allows us to set up controlled and accurate experiments for measuring \goalone and \goaltwo, as S2E typically succeeds in either goal in about one minute for each of the 72 functions. This makes measuring obfuscation overheads feasible, with a 1-hour budget per experiment sufficient to capture a slowdown of $\mathord{\sim}50\textrm{x}$ or higher. With 15 configurations and 2 goals, the tests took $>2000$ CPU hours.

% to skip unnecessary work for S2E
In light of all the considerations made in \Cref{ss:eval-deobfuscation}, we use a \puh{\ropobf{k}} setup with \predone and \predthree enabled (\predtwo and gadget confusion \puh{are disabled as they do not affect \dse}), with the same $\{s, n, p\}$ settings mentioned there for \predone, and with \predthree instantiated in its first variant and applied at different fractions $k$ of program points (\Cref{tab:variants}). As state exploration strategy for S2E we use class-uniform path analysis~\cite{cupa-candea} as it consistently yielded the best results across all ROP and VM configurations: its state grouping seems to work effectively for reducing bias towards picking hot spots involved in path explosion, which could be the case with \predthree instances under other strategies.

\begin{table}[t!]
\begin{footnotesize}
\begin{center}
\resizebox{0.38\textwidth}{!} {% was .435
\begin{tabular}{|l|c|c|c|}\hline
 {\sc Obfuscation} & \multicolumn{2}{c|}{{\sc Secret Finding}} & {\sc Code Coverage} \\ \cline{2-4}
 {\sc Configuration} & {\sc Found} & {\sc Avg Time} & {\sc 100\% Points} \\ \hline\hline
 {\sc Native} & 70/72 & 65.2s & 72/72 \\ \hline\hline
  \ropobf{0.05} & 19/72 & 907.9s & 34/72 \\
 \ropobf{0.25} & 10/72 & 568.4s & 11/72 \\
 \ropobf{0.50} & 9/72 & 884.0s & 9/72 \\
 \ropobf{0.75} & 5/72 & 775.3s & 7/72 \\
 \ropobf{1.00} & 1/72 & 3028.7s & 6/72 \\ \hline\hline
 \impvm{1}{all} & 61/72 & 85.8s & 68/72 \\
 \normvm{2} & 62/72 & 71.6s & 67/72 \\ 
 \impvm{2}{first} & 62/72 & 100.4s & 66/72 \\ 
 \impvm{2}{last} & 61/72 & 104.1s & 65/72 \\ 
 \impvm{2}{all} & 62/72 & 160.6s & 64/72 \\
 \normvm{3} & 62/72 & 119.2s & 69/72 \\ 
 \impvm{3}{first} & 54/72 & 899.2s & 56/72 \\ 
 \impvm{3}{last} & 62/72 & 240.3s & 61/72 \\ 
 \impvm{3}{all} & 0/72 & - & 0/72 \\ \hline

 %{\sc 1VM-PcUpd$_{all}$ + ROP$_{0.25}$} & 0/72 & - & 0/72 \\
 %{\sc 2VM-PcUpd$_{last}$ + ROP$_{0.25}$} & 14/72 & 784.7s & 15/72 \\ \hline
\end{tabular}
}
\end{center}
\end{footnotesize}
\vspace{-2.5mm}
\caption{Successful \scz{attacks} in the 1h-budget per program.\label{tab:obfuscation}}
\vspace{-3mm}
\end{table}

\begin{figure}[t!]
\centering
\includegraphics[width=0.8\linewidth]{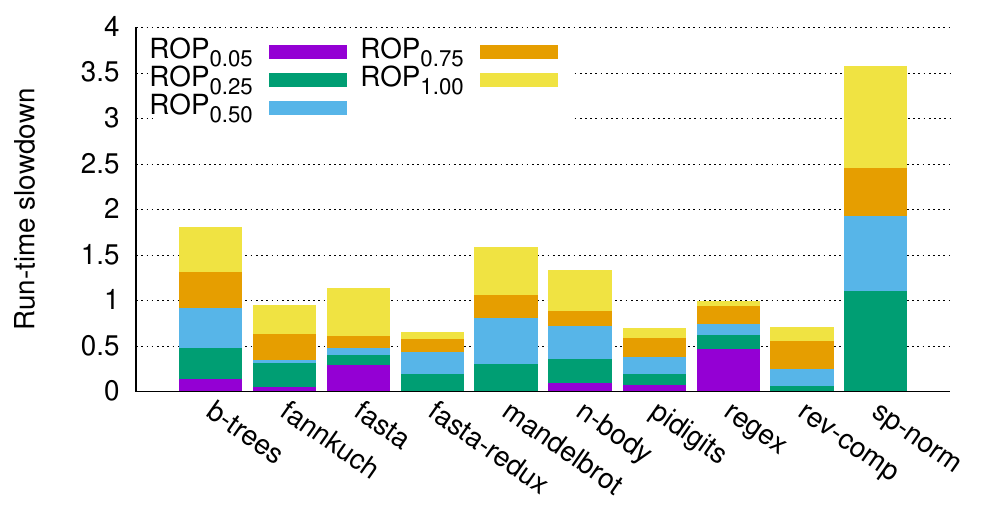}
\vspace{-3mm}
\caption{Run-time overhead for {\tt clbg} benchmarks of different \ropobf{k} settings with \impvm{2}{last} used as baseline.\label{fig:runtime-overhead}}
\vspace{-1.75mm}
\end{figure}

\subsubsection{\bf\em Secret Finding} Column two and three of \Cref{tab:obfuscation} summarize the results for the differently obfuscated configurations: for each class we report for how many functions S2E found the secret, and the average time \puh{for} successful attempts. For 2 of the 72 non-obfuscated functions S2E failed also with a 3-hour budget, likely due to excessively complex path constraints.

%\footnotetext{We do not report data for \normvm{1} and \ropobf{k=0} programs since S2E breaks them with no appreciable \scz{slowdown} w.r.t. their non-obfuscated counterparts.}

Coherently with insights from previous works~\cite{salawan,bardin-acsac19}, applying one or two layers of VM obfuscation does not prevent S2E from solving the majority of the secrets (the same sets of 61-62 functions over 72) even when using implicit VPC loads\footnotemark{}, with average overheads as high as $1.6\textrm{x}$ when applied to either the inner or the outer VPC, and $2.46\textrm{x}$ when to both. For \normvm{3} \scz{implicit VPC} loads are significantly more effective in slowing down S2E when applied on the innermost VPC than on the outermost one, while when used at all the three layers S2E found zero secrets within the 1-hour budget.

\begingroup
\setlength{\thickmuskip}{2.5mu}
The fraction of successful attacks to \ropobf{k} is lower than for VM configurations already for $k=0.05$, except for \impvm{3}{all} that however, as we see \puh{in \Cref{sss:runtime-overhead}}, may bring a destructive impact on program running time. The fraction of \ropobf{k}-protected functions that S2E can crack decreases with $k$: while we cannot compare average times computed for different sets, individual figures reveal that S2E can crack only the simpler functions as $k$ increases, with a higher processing time compared to when they were cracked for a smaller $k$.
\endgroup

\begingroup
\setlength{\thickmuskip}{2.5mu}
\subsubsection{\bf\em Code Coverage} The last column of \Cref{tab:obfuscation} lists for how many functions S2E covered all the CFG split and join points annotated by Tigress and reachable in the native counterparts (as the functions are relatively small, we consider coverage an ``all or nothing'' goal like in~\cite{bardin-acsac19, banescu-acsac16}). As seen in \Cref{ss:principles}, we recall that secret finding \puh{may} not require full coverage (neither achieving \goaltwo is sufficient for \goalone). For most VM configurations, the functions for which S2E fully explored the original CFG are slightly more than those for which it recovered the secret. $\textrm{ROP}_k$ already for $k=0.05$ impedes achieving \goaltwo for nearly half of the functions, and leaves only a handful (6-11) within the reach of S2E for higher $k$ values.
\endgroup

\subsection{Deployability}
\label{ss:obfuscation-eval-perf}
%  scenarios
To conclude our evaluation, we investigate how our methods can cope with real-world code bases in three respects: efficacy of the rewriting, run-time overhead for CPU-intensive code, and an obfuscation case study on a popular encoding function.

% whose size is below the 22 bytes required for our pivoting sequence
\subsubsection{\bf\em Coverage}
\label{sss:eval-coverage}
We start by assessing how our implementation can handle the code base of the {\tt coreutils} (v8.28, compiled with gcc 6.3.0 {\tt -O1}). Popular in software testing, this suite is a suitable benchmark thanks to its heterogeneous code patterns. Using symbol and size information, we identify 1354 unique functions across its corpus of 107 programs. We skip the 119 functions shorter than the 22 bytes the pivoting sequence requires\ifonlinereport\footnotemark{}\else\ifdefined\submittedpaper\footnotemark{}\fi\fi ~(\Cref{sss:materialization}). Our rewriter could transform 1175 over 1235 remaining  functions ($95.1\%$, or a $0.801$ fraction if normalized by size). 40 failures happened during register allocation as one spilling slot was not enough to cope with high pressure (\Cref{ss:design-discussion}), 19 \puh{for} code like {\tt push rsp} and {\tt push qword [rsp + imm]} that the translation step does not handle yet (\Cref{sss:translation}), and 1 for failed CFG reconstruction.

As informal validation of functional correctness, we run the test suite of the {\tt coreutils} over the obfuscated program instances, obtaining no mismatches in the output they compute.
%Two tests that use valgrind~\cite{valgrind} for memory sanitization raised false alerts for uninitialized stack reads, which is indeed the way valgrind would see the fetching of a ROP chain element from memory.

\begingroup
\setlength{\thickmuskip}{2.5mu}
%Computer Benchmarks Game
\subsubsection{\bf\em Overhead}
\label{sss:runtime-overhead}
%While heavy-duty obfuscations commonly target one-off or infrequent computations, we find it interesting to examine a worst-case scenario for run-time overhead by taking into account}{
% sought fa occupare rigo extra :)
\looseness=-1
Albeit a common assumption is that heavy-duty obfuscation target one-off or infrequent computations, we also seek to study performance overhead aspects. We consider the {\tt clbg} suite~\cite{computer-bench-game} used in compiler research to benchmark the effects of code transformations (e.g.,~\cite{castanos-oopsla12,daniele-cgo16}). As a reference we consider \impvm{2}{last} as it was the fastest configuration for double virtualization with implicit VPC loads (\normvm{1} is too easy to circumvent, and \normvm{3} brings prohibitive overheads, i.e., over 5-6 orders of magnitude in our tests). % as in our tests was

\Cref{fig:runtime-overhead} uses a stacked barchart layout to present slowdowns for \ropobf{k}, as its overhead can only grow with $k$.
%As the overhead \marco{}{for \ropobf{k}} can only grow with $k$ \marco{because we add more P3 instances}{(i.e., more P3 instances inserted)}, we use a stacked barchart layout \marco{for presenting}{to present} slowdowns in \Cref{fig:runtime-overhead}.
With the exception of {\tt sp-norm} \scz{that sees repeated pivoting events from a ROP tight loop calling a short-lived ROP subroutine}, \ropobf{k} is consistently faster than \impvm{2}{last} for $k\le0.5$, and no slower than $1.81\textrm{x}$ \scz{({\tt b-trees} that repeatedly calls {\tt malloc} and {\tt free})} when in the most expensive setting $k=1.00$. %$\textrm{ROP}_k$ thus seems to bring better path protection (\Cref{tab:obfuscation}) than VM configurations with even fewer resources. % sono bench diversi
%where some chain manipulations pollute the locality of matrix accesses
\endgroup

\subsubsection{\bf\em Case Study} Finally, we study resilience and slowdowns of selected obfuscation configurations on the reference implementation of the popular base64 encoding algorithm~\cite{base64}.
base64 features byte manipulations and table lookups relevant for transformation code of variable complexity that users may wish to obfuscate. An important consideration is that in the presence of table lookups, using concrete values for input-dependent pointers is no longer effective (but even counter-productive) for \dse to explore relevant states. We thus opt for the per-page theory-of-arrays (\cite{symbolic-memory,survey-baldoni}) memory model of S2E. This choice allows S2E to recover a 6-byte input in about 102 seconds for the original implementation, \scz{180 for \impvm{2}{last}, 281 for \impvm{2}{all},} and 1622 for \impvm{3}{last}.

\begingroup
\setlength{\thickmuskip}{2.5mu}
A budget of 8 hours was not sufficient for \impvm{3}{all}, as well as for \ropobf{k} already for $k=0$ \puh{(when} only \predone~\puh{is enabled)}. As anticipated in \Cref{ss:discussion-predicates}, the aliasing from \predone on RSP updates can impact the handling of memory \puh{in} \dse executors in ways that the synthetic functions of \Cref{ss:obfuscation-eval-robustness} did not \puh{(as they do not use table lookups)}. As for code slowdowns, $\textrm{ROP}_k$ seems to bring rather tolerable execution times: \scz{for a rough comparison}, execution takes 0.299ms for \ropobf{0.25} and 1.791ms for \ropobf{1.00}, while for VM settings we measured \scz{1.63ms for \impvm{2}{last}, 347ms for \impvm{2}{all}, 668ms for \impvm{3}{last} and 2211s for the unpractical \impvm{3}{all}}.
\endgroup

\ifonlinereport
\footnotetext[6]{We do not report data for \normvm{1} and \ropobf{k=0} programs since S2E breaks them with no appreciable \scz{slowdown} w.r.t. their non-obfuscated counterparts.}
\footnotetext[7]{While we could have added a trampoline to some code cavity large enough to hold it, these functions appear to be stubs of unappealing complexity.}
\else
\ifdefined\submittedpaper
\footnotetext{\cready{}{While we could have added a trampoline to some code cavity large enough to hold it, these functions appear to be stubs of unappealing complexity.}}
\fi
\fi

\section{Concluding Remarks}
\label{se:conclusions}
% qualitative?
Adding to the appealing properties of ROP against reverse engineering that we discussed throughout the paper, the experimental results lead us to believe that our approach can:
\begin{enumerate}
\item hinder many popular deobfuscation approaches, as well as symbiotic combinations aimed at ameliorating scalability;
\item significantly increase the resources needed by automated techniques that remain viable, with slowdowns $\mathrel{\geq}50\textrm{x}$ for the vast majority of the 72 targets for both end goals \goalboth;
\item bring multiple configuration opportunities for resilience (and overhead) goals to the program protection landscape. 
\end{enumerate}

While obfuscation research is yet to declare a clear winner and automated attacks keep evolving, our technique is also orthogonal to most other code obfuscations, meaning it can be applied on top of already obfuscated code (\Cref{ss:design-discussion}). \puh{We have} followed established practices~\cite{TutorialBanescu} of analyzing our obfuscation individually and on function units, \puh{yet} in future work we would like to expand both points: namely, studying mutually reinforcing combinations with other obfuscations, and applying ROP rewriting inter-procedurally, removing the stack-switching step during transfers between ROP functions, since our design allows that. % so to hide calls. %) to hide also calls.
Finally, to optimize composition of symbolic registers when instantiating \predthree, we may look at def-use chains as suggested by~\cite{bardin-acsac19} for {\sc For} cases, exploring analyses like~\cite{value-set-analysis} necessary to obtain the required information.%, while when the subject of obfuscation is source code extending standard compiler analyses may suffice

%% file: discussion.tex
% !TEX root = main.tex
\iffalse
\section{Scratchpad}
Countermeasures for an adversary tailored to our technique?
Interplay with ROP defenses
What happens with the deobfuscation techniques that we do not evaluate against (say backward slicing - no available implementation is a factor)
We can still combine our technique with other obfuscations (inner layers) => protecting data vs protecting instructions
Our approach applied to other code reuse paradigms

We leave to future work exploring how our technique fares when combined with other obfuscation techniques. In this regard we conform to the trend in the code obfuscation literature, where techniques are evaluated in isolation and only recently a few works have advocated for exploratory studies on the effectiveness of combined transformations (Banescu, then?).

\fi

%% file: conclusion.tex
% !TEX root = main.tex

%\section{Conclusion}

%% file: appendix.tex
% !TEX root = ms.tex

%\newpage
\appendices

\section{Additional Material}
\label{apx:a}

We report practical details of the rewriter implementation, and additional settings and findings from the evaluation.

\begin{table*}[t!]
\begingroup
\setlength{\thickmuskip}{1.5mu}
\begin{small}
\begin{center}
\resizebox{0.99\textwidth}{!} {
\begin{tabular}{|l||c|a|b|d|a|b|d|a|b|d|a|b|d|a|b|d|a|b|d|}\hline
\multirow{2}{*}{{\sc Benchmark}} & \multirow{2}{*}{{\sc N}} & \multicolumn{3}{c|}{ROP$_{0.00}$} & \multicolumn{3}{c|}{ROP$_{0.05}$} & \multicolumn{3}{c|}{ROP$_{0.25}$} & \multicolumn{3}{c|}{ROP$_{0.50}$} & \multicolumn{3}{c|}{ROP$_{0.75}$} & \multicolumn{3}{c|}{ROP$_{1.00}$} \\\hhline{|~|~|------------------|}
 & & A & B & C & A & B & C & A & B & C & A & B & C & A & B & C & A & B & C \\\hline\hline
{\sc b-trees}& 170 & 583 & 144 & 3.43 & 966 & 322 & 5.68 & 1513 & 555 & 8.90 & 2090 & 719 & 12.29 & 2734 & 856 & 16.08 & 3411 & 940 & 20.06 \\\hline
{\sc fannkuch} & 89 & 193 & 104 & 2.17 & 471 & 199 & 5.29 & 798 & 377 & 8.97 & 1085 & 491 & 12.19 & 1310 & 589 & 14.72 & 1608 & 644 & 18.07 \\\hline
{\sc fasta} & 256 & 733 & 226 & 2.86 & 1372 & 431 & 5.36 & 2135 & 733 & 8.34 & 3015 & 919 & 11.78 & 3830 & 1025 & 14.96 & 4717 & 1124 & 18.43 \\\hline
{\sc fasta-redux} & 263 & 670 & 246 & 2.55 & 1313 & 499 & 4.99 & 1775 & 676 & 6.75 & 2842 & 950 & 10.81 & 3914 & 1105 & 14.88 & 4727 & 1163 & 17.97 \\\hline
{\sc mandelbrot} & 135 & 323 & 143 & 2.39 & 793 & 316 & 5.87 & 1047 & 419 & 7.76 & 1598 & 539 & 11.84 & 2341 & 642 & 17.34 & 2666 & 659 & 19.75 \\\hline
{\sc n-body} & 288 & 680 & 260 & 2.36 & 1443 & 535 & 5.01 & 2026 & 747 & 7.03 & 3259 & 1003 & 11.32 & 4444 & 1137 & 15.43 & 5487 & 1220 & 19.05 \\\hline
{\sc pidigits} & 144 & 462 & 120 & 3.21 & 814 & 285 & 5.65 & 1189 & 457 & 8.26 & 1769 & 621 & 12.28 & 2211 & 710 & 15.35 & 2883 & 803 & 20.02 \\\hline
{\sc regex-redux} & 162 & 522 & 147 & 3.22 & 869 & 285 & 5.36 & 1437 & 546 & 8.87 & 1946 & 713 & 12.01 & 2539 & 844 & 15.67 & 3268 & 935 & 20.17 \\\hline
{\sc rev-comp} & 176 & 558 & 174 & 3.17 & 1265 & 364 & 7.19 & 1770 & 605 & 10.06 & 2322 & 720 & 13.19 & 2810 & 838 & 15.97 & 3367 & 924 & 19.13 \\\hline
{\sc sp-norm} & 115 & 329 & 119 & 2.86 & 529 & 195 & 4.60 & 845 & 381 & 7.35 & 1361 & 554 & 11.83 & 1761 & 660 & 15.31 & 2234 & 737 & 19.43 \\\hline\hline
{\sc Avg / Geo. Mean} & 179.80 & 505.30 & 168.30 & 2.79 & 983.50 & 343.10 & 5.46 & 1453.50 & 549.60 & 8.17 & 2128.70 & 722.90 & 11.94 & 2789.40 & 840.60 & 15.55 & 3436.80 & 914.90 & 19.19 \\\hline
\end{tabular}
}
\end{center}
\end{small}
\caption{N$:\mathrel{=}$ number of program points (instructions), A$:\mathrel{=}$ total number of gadgets used across all ROP chains; B$:\mathrel{=}$ number of unique gadgets used across all ROP chains; C$:\mathrel{=}$ average number of gadgets used per program point.\label{tab:obfuscation-stats}}
%\vspace{-4mm}
\endgroup
\end{table*}

\subsection*{Implementation Aspects}
%\subsection{Implementing Switch Tables in ROP}
\subsubsection*{Switch Tables} As compilers recur to indirect jumps to efficiently implement switch constructs, we use CFG reconstruction heuristics to reveal possible targets, and then use the original target calculation sequence used for the dispatching to our advantage. Instead of jumping to the location corresponding to the desired code block in the original program, at rewriting time we store at such location the RSP displacement in the chain for the corresponding code block, then during execution we use the address computed by the dispatching sequence to read the correct offset. Thus when lowering, e.g., a {\tt jmp reg} the rewriter can combine gadgets to achieve:
\begin{verbatim}
    movsx reg, byte ptr [reg]  ## offset for RSP
    shl reg, 0x3
    add rsp, reg
\end{verbatim}
where {\tt reg} contains the jump target address for the original program. In the end, the main difference in our ROP encoding between a direct jump and an indirect jump is that the former retrieves the offset to add to RSP from the chain (\Cref{sss:crafting}), while the latter retrieves it from the location of the original code block. We use offsets of 1, 2, o 4 bytes depending on the position of the desired block in the chain from the jump site (the example uses a 1-byte value).

\iffalse
We report an example ROP implementation for an indirect jump in the form of {\tt jmp reg}. Each line represents a possible gadget ending with a {\tt ret}:

\begin{verbatim}
    movsx reg, byte ptr [reg]
    shl reg, 0x3
    add rsp, reg
\end{verbatim}

The {\it CFG reconstruction} module provides the list of all the possible jump targets for each indirect jump.
The translation strategy ensures that the register holding the target of the jump is preserved, so at runtime the gadgets mirroring the {\tt jmp} instruction will have the target information available. We build at translation time a reverse jump table, where for each possible address destination of a jump we store the respective offset needed to update {\tt rsp} to perform the jump as shown for direct jumps. The indirect jump gadgets, therefore, will only have to access the target address, held at runtime in the destination register, for the {\tt jmp} to retrieve the correct offset to apply to {\tt rsp}. Each offset is encoded as a signed {\tt byte}/{\tt word}/{\tt dword} as needed.
\fi

%\vspace{-2pt}
\subsubsection*{From Native to ROP and Back}
Upon generation of a ROP chain the rewriter replaces the original function implementation in the program with a stub that switches the stack and activates the chain (\Cref{sss:materialization}). The pivoting stub performs three steps: (a) it reserves a new entry for storing {\tt other\_rsp} in the stack-switching array {\tt ss}, (b) it saves RSP in {\tt other\_rsp}, (c) it loads the address of the chain ({\tt chain\_address}) for the function to execute into RSP and starts the chain execution. A pivoting stub can be implemented in 22 bytes as:
\begin{verbatim}
    push ss                     
    pop rax                     
    add qword ptr [rax], 0x8    
    add rax, qword ptr [rax]    ## step (a) ends
    mov qword ptr [rax], rsp    ## step (b)
    push chain_address          ## step (c)  
    pop rsp                     
    ret                         
\end{verbatim}
This code is optimized to use only a single caller-save register (e.g., {\tt rax}) and uses {\tt push}-{\tt pop} sequences in place of {\tt mov} instructions to minimize the number of bytes required to encode the sequence.

When a chain reaches its epilogue and a native function has to be resumed, a symmetric {\em unpivoting} sequence takes care of removing the entry created for storing {\tt other\_rsp} for the active chain and restores the native stack pointer into RSP.

The implementation should realize e.g.:
\begin{verbatim}
    pop r11                     ## ss
    sub qword ptr [r11], 0x8 
    add r11, qword ptr [r11]
    add r11, 0x8
    mov rsp, qword ptr [r11]
\end{verbatim}
Notice that only register {\tt r11} will be clobbered by this sequence. Also, while the pivoting sequence is made of native instructions, the sequence above will be realized by gadgets.

\subsubsection*{Tail Jumps}
To handle optimized tail jump instances the rewriter uses an unpivoting variant that, instead of returning to the calling function, jumps to the target of the tail jump:

\begin{verbatim}
    pop r11                     ## ss
    sub qword ptr [r11], 0x8 
    add r11, qword ptr [r11]
    add r11, 0x8
    pop rax                     ## jmp target
    mov rsp, qword ptr [r11]; jmp rax
\end{verbatim}

The last line represents a JOP gadget.

\smallskip
\subsection*{Evaluation Additions}

\subsubsection*{Functions for Obfuscation Resilience Experiments}
To produce the 72 hash functions for the \goalone tests of \Cref{ss:obfuscation-eval-robustness} we used the {\it RandomFuns} feature of Tigress with a command line:

\iffalse
Each random generated function for the {\it secret finding} evaluation was produced by the {\it RandomFuns} transformation of the Tigress obfuscator, using the following command, with {\tt body\_structure}'s possible values listed in Table~\ref{tab:body}, {\tt seed} $\in \{1,2,3\}$ and {\tt data\_type} $\in$ {\tt \{char, short, int, long\}}:
\fi

\begin{verbatim}
tigress --Verbosity=1 --Seed={seed} 
    --Environment=x86_64:Linux:Gcc:6.3.0 
    --Transform=RandomFuns 
        --RandomFunsName=target 
        --RandomFunsTrace=0 
        --RandomFunsType={data_type} 
        --RandomFunsInputSize=1 
        --RandomFunsLocalStaticStateSize=1 
        --RandomFunsGlobalStaticStateSize=0 
        --RandomFunsLocalDynamicStateSize=1 
        --RandomFunsGlobalDynamicStateSize=0 
        --RandomFunsBoolSize=3 
        --RandomFunsLoopSize=25 
        --RandomFunsCodeSize=1000 
        --RandomFunsOutputSize=1 
        --RandomFunsControlStructures={control}
        --RandomFunsPointTest=true 
    --out={output_file} {base_path}/empty.c
\end{verbatim}
and different combinations for the parameters {\tt control}, {\tt seed}, and {\tt data\_type}. Table~\ref{tab:body} lists the values we used for {\tt control}; {\tt seed} was from $\{1, 2, 3\}$, while {\tt data\_type} was {\tt char}, {\tt short}, {\tt int}, or {\tt long}.

For the {\it code coverage} scenario we used the same sets of parameters and altered the command in the following way: we set {\tt Random} {\tt FunsPointTest} to {\tt false} to disable the secret value checking step, and {\tt RandomFunsTrace} to {\tt 2} to annotate CFG split and join points.% with probes that assist us when measuring code coverage points.

%%% KTM

%\subsection{Additional Experiments}

\medskip
\subsubsection*{VM Obfuscation}
To apply during our experiments one or more layers of VM obfuscation (with or without implicit VPC loads) to a piece of code we used Tigress with the following parameters:

\begin{verbatim}
tigress --Environment=x86_64:Linux:Gcc:6.3.0 
    --Transform=InitOpaque --Functions=main 
        --Transform=InitImplicitFlow --Functions=main 
        --InitImplicitFlowHandlerCount=0 
        --InitImplicitFlowKinds=counter_int, \
        counter_float,bitcopy_unrolled,bitcopy_loop 
    [transformations]
    --out={output_file} {input_file}
\end{verbatim}
Where {\tt [transformations]} is obtained by adding for each VM layer on {\tt function} the following parameters:
\begin{verbatim}
    --Transform=Virtualize 
        --VirtualizeDispatch={[call, switch]} 
        --VirtualizeImplicitFlowPC={[PCUpdate,none]} 
        --VirtualizeOpaqueStructs=array 
        --Functions={function}
\end{verbatim}

Following the documentation of Tigress, we alternate the {\em call} and {\em switch} methods for VPC dispatching across nested virtualization layers. In our tests with S2E, however, the engine did not seem evidently affected by the choice of a particular dispatching method over another (including the additional schemes supported by Tigress to this end).\vfill

\begin{table}[t!]
\begingroup
\setlength{\thickmuskip}{1.5mu}
\begin{small}
\begin{center}
\resizebox{0.45\textwidth}{!} {
\begin{tabular}{|l|p{0.7cm}|p{0.7cm}|p{0.7cm}|}\hline
 {\bf\tt RandomFunsControlStructures} {\bf Parameter Value} & {\bf Ctrl-flow depth} & {\bf Num. of {\bf\it if}\bf-stmts} & {\bf Num. of Loops} \\ \hline\hline
(if (bb 4) (bb 4)) & 1 & 1 & 0 \\ \hline
(for (if (bb 4) (bb 4))) & 2 & 1 & 1 \\ \hline 
(for (for (bb 4))) & 2 & 0 & 2 \\ \hline 
(for (for (if (bb 4) (bb 4)))) & 3 & 1 & 2 \\ \hline 
(for (if (if (bb 4) (bb 4)) (if (bb 4) (bb 4)))) & 3 & 3 & 1 \\ \hline
(if (if (if (bb 4) (bb 4)) (if (bb 4) (bb 4))) (if (bb 4) (bb 4))) & 3 & 5 & 0 \\ \hline
\end{tabular}
}
\end{center}
\end{small}
\vspace{-0.5em}
\caption{Values used for the {\tt RandomFuns ControlStructures} parameter in Tigress to generate the 72 functions for \Cref{ss:obfuscation-eval-robustness}.\label{tab:body}}
\vspace{-1mm}
%\vspace{-1em}
\endgroup
\end{table}

%\balance
\subsubsection*{Additional Deployability Experiments}
Table~\ref{tab:obfuscation-stats} reports several statistics collected by the rewriter during the translation of the 10 benchmarks from the {\tt shootout} suite considered in \Cref{ss:obfuscation-eval-perf}. In particular, the table provides for different settings of \ropobf{k} the following kinds of information: the number of program points (i.e., instructions) N that were obfuscated (not affected by $k$), the total number A of gadgets used by the chains from all the obfuscated functions, the number B of unique gadgets used across all chains, and the average number C of gadgets used per obfuscated program point. 

A, B, and C increase as we add more instance of P3 with higher $k$ values. \scz{The increase in the number B of used unique gadgets shall be interpreted as an indication that the rewriter does not use a fixed set of gadgets to instantiate P3 (but can draw from multiple equivalent versions of a desired gadget functionality, bringing diversity) and that possibly different dead and symbolic registers get involved at different program points. At the same time, the rewriter performs gadget reuse across different chains: for instance, for $k = 1.00$ the rewriter uses the largest number of gadgets per obfuscated instruction (geometric mean of $19.19$), but the ratio between the average values for A and B is $\mathord{\sim}3.75$, showing that a gadget is on average reused almost four times across all the chains.}
%\vfill

%\pagebreak